# Environmental effects in the interaction and merging of galaxies in zCOSMOS[1]

P. Kampczyk[2], S. J. Lilly[2], L. de Ravel[3, 4], O. Le Fèvre[3], M. Bolzonella[5], C. M. Carollo[2], C. Diener[2], C. Knobel[2], K. Kovač[2, 6], C. Maier[2], A. Renzini[7], M. T. Sargent[8], D. Vergani[5], U. Abbas[9], S. Bardelli[5], A. Bongiorno[10], R. Bordoloi[2], K. Caputi[11], T. Contini[12, 13], G. Coppa[5, 10, 14], O. Cucciati[3], S. de la Torre[15], P. Franzetti[16], B. Garilli[16], A. Iovino[15], J.-P. Kneib[3], A. M. Koekemoer[17], F. Lamareille[12, 13], J.-F. Le Borgne[12, 13], V. Le Brun[3], A. Leauthaud[9], V. Mainieri[18], M. Mignoli[5], R. Pello[12, 13], Y. Peng[2], E. Perez Montero[12, 13, 19], E. Ricciardelli[20], M. Scodeggio[17], J. D. Silverman[21], M. Tanaka[21], L. Tasca[16], L. Tresse[3], G. Zamorani[5], E. Zucca[5], D. Bottini[16], A. Cappi[5], P. Cassata[22], A. Cimatti[5], M. Fumana[16], L. Guzzo[15], J. Kartaltepe[23], C. Marinoni[24], H. J. McCracken[25], P. Memeo[16], B. Meneux[10, 26], P. Oesch[2], C. Porciani[2], L. Pozzetti[5], R. Scaramella[27]

[1] Based on observations obtained at the European Southern Observatory (ESO) Very Large Telescope (VLT), Paranal, Chile, as part of the Large Program 175.A-0839 (the zCOSMOS Spectroscopic Redshift Survey)
[2] Institute of Astronomy, ETH Zürich, CH-8093, Zürich, Switzerland; kampczyk@phys.ethz.ch
[3] Laboratoire d'Astrophysique de Marseille, UMR 6110 CNRS-Universite de Provence, BP8, F-13376 Marseille Cedex 12, France
[4] Institute for Astronomy, University of Edinburgh, Royal Observatory, Edinburgh, EH93HJ, UK
[5] INAF Osservatorio Astronomico di Bologna, via Ranzani 1, I-40127, Bologna, Italy
[6] Max–Planck–Institut fuer Astrophysik, Karl–Schwarzschild–Str. 1, D-85748 Garching, Germany
[7] INAF - Osservatorio Astronomico di Padova, Vicolo dell'Osservatorio 5, 35122, Padova, Italy
[8] Laboratoire AIM, CEA/DSM-CNRS-Universite Paris Diderot, Irfu/Service dAstrophysique, CEA-Saclay, Orme des Merisiers, 91191 Gif-sur-Yvette Cedex, France
[9] Berkeley Lab & Berkeley Center for Cosmological Physics, University of California, Lawrence Berkeley National Lab., 1 cyclotron road, MS 50-5005, Berkeley, CA, USA
[10] Max-Planck-Institut für Extraterrestrische Physik, D-84571, Garching, Germany
[11] SUPA, Institute for Astronomy, University of Edinburgh, Royal Observatory, Edinburgh - EH9 3HJ, UK
[12] Institut de Recherche en Astrophysique et Planétologie, CNRS, 14, avenue Edouard Belin, F-31400 Toulouse, France
[13] IRAP, Université de Toulouse, UPS-OMP, Toulouse, France
[14] Dipartimento di Astronomia, Universita di Bologna, Bologna, Italy
[15] INAF Osservatorio Astronomico di Brera, Milan, Italy
[16] INAF - IASF Milano, Milan, Italy
[17] Space Telescope Science Institute, 3700 San Martin Drive, Baltimore, MD 21218-2410, USA
[18] European Southern Observatory, Karl-Schwarzschild-Strasse 2, Garching, D-85748, Germany
[19] Instituto de Astrofísica de Andalucía, CSIC, Apartado de correos 3004, 18080 Granada, Spain
[20] Dipartimento di Astronomia, Universita di Padova, Padova, Italy
[21] Institute for the Physics and Mathematics of the Universe (IPMU), University of Tokyo, Kashiwanoha 5-1-5, Kashiwa, Chiba 277-8568, Japan
[22] Dept. of Astronomy, University of Massachusetts at Amherst, USA
[23] National Optical Astronomy Observatory, 950 North Cherry Avenue, Tucson, AZ 85719, USA
[24] Centre de Physique Theorique, Marseille, Marseille, France
[25] Institut d'Astrophysique de Paris, UMR7095CNRS, Universite Pierre et Marie Curie, 98 bis Boulevard Arago, F-75014 Paris, France
[26] Universitäts-Sternwarte, Scheinerstrasse1, Munich D-81679, Germany
[27] INAF - Osservatorio Astronomico di Roma, Osservatorio Astronomico di Roma, Via di Frascati 33, 00040 Monte Porzio Catone, Italy

**Abstract**

We analyze the environments and galactic properties (morphologies and star-formation histories) of a sample of 153 close kinematic pairs in the redshift range $0.2 < z < 1$ identified in the zCOSMOS-bright 10k spectroscopic sample of galaxies. Correcting for projection effects, the fraction of close kinematic pairs is three times higher in the top density quartile than in the lowest one. This translates to a three times higher merger rate because the merger timescales are shown, from mock catalogues based on the Millennium simulation, to be largely independent of environment once the same corrections for projection is applied. We then examine the morphologies and stellar populations of galaxies in the pairs, comparing them to control samples that are carefully matched in environment so as to remove as much as possible the well-known effects of environment on the properties of the parent population of galaxies. Once the environment is properly taken into account in this way, we find that the early-late morphology mix is the same as for the parent population, but that the fraction of irregular galaxies is boosted by 50-75%, with a disproportionate increase in the number of irregular-irregular pairs (factor of 4-8 times), due to the disturbance of disk galaxies. Future dry-mergers, involving elliptical galaxies comprise less than 5% of all close kinematic pairs. In the closest pairs, there is a boost in the specific star-formation rates of star-forming galaxies of a factor of 2-4, and there is also evidence for an increased incidence of post star-burst galaxies. Although significant for the galaxies involved, the "excess" star-formation associated with pairs represents only about 5% of the integrated star-formation activity in the parent sample. Although most pair galaxies are in dense environments, the effects of interaction appear to be largest in the lower density environments. By preferentially bringing more pairs into the sample in lower density environments, this could dilute the dependence of pair fraction on environment, and may introduce other biases into the observational study of mergers especially those based on morphological criteria.

Subject headings: galaxies: evolution — galaxies: interactions — galaxies: general — galaxies: starburst

## 1. Introduction

It has been clear for many years that the merging of galaxies is an important phenomenon in the extragalactic Universe. However its overall role in the assembly of the stellar mass in today's massive galaxies and its role in controlling or modifying the star-formation history of galaxies, e.g. through "quenching", is still unclear.

In the standard scenario for formation of structures in the Universe, cold dark matter (CDM) haloes hierarchically merge to form larger and larger structures. This process has been addressed in a number of theoretical studies and simulations which yield a prediction of a steeply increasing merger rate of CDM haloes with redshift for major mergers, defined as a mass ratio of 1:4 or less. With evolution parameterized as $(1+z)^m$ the exponent $m$ is believed to be in a relatively tight range $2.5 < m < 3.5$ (Governato et al. 1999; Gottlöber et al. 2001). The fate of the baryons, whether gaseous or stellar, within these merging CDM haloes is more complicated. This is due to the complexity of gas physics and of the relevant dynamical processes. These, together with a number of potential observational biases, complicates the relatively simple picture traced by the dark matter.

Extensive studies in the last two decades have tried to address the importance of galaxy mergers in shaping the galaxy population. Most studies look at the evolution of the galaxy merger fraction, i.e. the fraction of galaxies seen to be undergoing a merger. This can be related to the merger *rates* via a rather uncertain and often underestimated time scale for the duration of the merger itself (see Kitzbichler and White 2008). Estimates of $m$ for systems identified as ongoing mergers or as close galaxy pairs soon to merge vary in the literature between $0 < m < 6$ (see e.g. Zepf & Koo 1989, Carlberg et al. 1994, Yee & Ellingson 1995, Neuschaefer et al. 1995, Carlberg et al. 2000, Le Fèvre et al. 2000, Patton et al. 2002, Lin et al. 2004, Conselice et al. 2003, Lavery et al. 2004, Kartaltepe et al. 2006, Kampczyk et al. 2007, Lin et al. 2008, Lotz et al. 2008, de Ravel et al. 2009, López-Sanjuan et al. 2009, Bridge et al. 2010). This variation likely reflects a number of different effects, including the use of different methods, different luminosity ranges of surveys, different selection criteria, various observational biases and different ways of handling incompleteness in galaxy surveys.

In Kampczyk et al. (2007) we attempted to derive a merger fraction in the COSMOS field at $z \sim 0.7$ based on purely morphological features and compared it in a consistent way with one derived from Sloan Digital Sky Survey (SDSS) at very low redshifts, taking into account the many observational biases by carefully simulating the appearance that galaxies at low redshifts would have if they were to be observed at high redshift. One of the interesting results of those studies was, that about 40% of the apparent morphologically-selected mergers identified on ACS images at $z = 0.7$ are likely to be spurious and arising only due to random projections with unrelated background or foreground objects, while only a minority of real SDSS mergers were recognized as such at $z \sim 0.7$, and none at $z \sim 1.2$. A sobering conclusion from that study was that morphological studies of merging rates are beset with observational difficulties.

The evolution and appearance of two merging galaxies is highly dependent on the gas content of the galaxies, which may in turn reflect a number of things including the morphological types, masses, redshifts, environments, and the previous star formation and merging histories of the individual galaxies. The most violent galaxy starbursts observed in the Universe are believed to be the result of mergers (e.g. Sanders & Mirabel 1996) and it is also known that mergers of gas-rich galaxies can trigger less extreme episodes of star formation (e.g. Barton et al. 2000; Lambas et al. 2003; Barton et al. 2007) and be responsible for changing the morphologies of galaxies resulting in spheroidal systems (Toomre & Toomre 1972). Merging of early type, gas-poor galaxies has been suggested as an important channel for producing massive elliptical galaxies (Khochfar et al., 2003; van Dokkum 2005; Bell et al., 2006; Di Matteo et al., 2007; Lin et al., 2008; Lin et al., 2010), however its importance is still under debate (see Scarlata et al., 2007a, De Propris et al. 2010, Peng et al. 2010).

The first observational evidence of enhanced star formation in paired galaxies goes back to Holmberg (1958). In a sample of 32 galaxy pairs, the colors of paired galaxies turned out to be closely correlated. Blue galaxies would have preferentially blue companions and similar correlation exists for the red galaxies. This phenomenon is widely called the "Holmberg effect". Merging galaxies are predicted to show enhancement of star formation in all of the phases of their interaction – starting from first passage up to the end of coalescence phase (e.g. Barnes & Hernquist 1996; di Matteo et al. 2007). Such enhancement has been observed in still well separated pairs of galaxies (e.g. Kennicutt et al. 1987; Lambas et al. 2003; Alonso et al. 2004; Patton et al. 2005; Barton et al. 2007; Smith et al. 2007; Woods & Geller 2007; Ellison et al. 2008; Lin et al. 2008; Li et al. 2008; Robaina et al. 2009, Ellison et al. 2010). On the other hand, pairs of spheroidal galaxies show few, if any, signatures of interactions, presumably due to their low gas content (Luo, Shu & Huang 2007; Park & Choi 2009; Rogers et al. 2009; Darg et al. 2010), making uniform and unbiased identification of a future mergers based on enhanced star formation or tidal features challenging, if not impossible.

The pre-encounter morphologies of galaxies will affect their interactions with other galaxies, and the form and outcome of the final mergers. From the discovery of the morphology-density relation (Dressler 1980) it has become clear that properties of galaxies, such as morphologies, colors and star formation rates, all depend on the environments that the galaxies reside in. Any subsample of galaxies that are drawn preferentially from specific environments may well therefore have a different distribution of properties to that of the overall galaxy population. It is therefore important to examine the environments of galaxies suspected of merging before looking for systematic changes induced by the merging, so that a valid comparison with a non-merging population can be made.

Recently a small number of studies have looked for a possible environmental dependence of merging. It has been shown locally, that the highest fractions of mergers are in intermediate to high-density regions (McIntosh et al., 2008; Darg et al., 2010; Perez et al., 2009; Ellison et al., 2010). This trend, while possibly somewhat weaker seems also to be present also out to $z \sim 1$ (Lin et al. 2010, de Ravel et al. 2011).

The current analysis is based on the first 10,000 "bright" galaxies with $I_{AB} <$ 22.5 of the zCOSMOS-bright survey – a major spectroscopic redshift survey (Lilly et al. 2007, 2009) in the COSMOS field (Scoville et al. 2007). With high angular and redshift completeness, and relatively precise velocity information, plus high resolution HST imaging (Koekemoer et al. 2007) and a wealth of ancillary photometric data (e.g. Capak et al. 2007), this survey is well suited for the study of close kinematic pairs. Furthermore, the same redshifts, supplemented with photometric redshifts for the galaxies not observed spectroscopically, have been used to reconstruct the three-dimensional density field (Kovač et al. 2010a), which has enabled several studies of the environmental drivers of galaxy evolution up to $z \sim 1$. The reader is referred to several studies utilizing derived over-densities in the zCOSMOS 10k bright sample. Bolzonella et al. (2009), Cucciati et al. (2010, in press), Tasca et al. (2009) and Zucca et al. (2009) analyze properties of the global galaxy samples i.e. mass and luminosity functions, color and morphological segregation as a function of environment. Caputi et al. (2009) and Silverman et al. (2009) establish environmental dependences for populations of 24 μm galaxies and AGN respectively. Iovino et al. (2010) and Kovač et al. (2010b) study morphologies and colors of group galaxies presented in Knobel et al. (2009), while comparing them to isolated galaxies. Peng et al. (2010) have presented a global model based on a comparison of zCOSMOS with the SDSS.

The goals of this paper are two-fold: First we wish to study the environmental dependence of the *close pair fractions*, taken as a proxy for the *merger rates*, as a function of environment. This is required as an input to understanding the build-up of stellar mass in different environments, as well as the possible or likely role of mergers in producing morphological transformations and/or the quenching of star-formation in galaxies in different environments (see e.g. Peng et al. 2010). This extends the earlier analysis of de Ravel et al. (2011), which primarily looked at the change in merger rate with epoch in the same zCOSMOS sample, and includes a careful consideration of possible biases with local density that may enter into the identification of pairs. Second, we wish to examine the properties of galaxies in close pairs relative to the general galaxy population to search for evidence of processes directly triggered by the interaction itself, e.g. possibly enhanced star-formation rates or morphological disturbances. However, as we show in this paper, such a comparison needs the careful control of the environmental dependences of galaxy properties, since many galaxy properties depend to a certain degree on the environment.

We use throughout the paper a concordance cosmology with $\Omega_\Lambda = 0.75$, $\Omega_M = 0.25$. For ease of comparison with the literature, we have adopted $H_0 = 100\ h$ kms$^{-1}$Mpc$^{-1}$, where $h = 0.7$. All magnitudes are quoted in the AB system (Oke 1974).

## 2. Selection of close kinematic pairs

### 2.1 zCOSMOS-bright 10k sample

This paper utilizes the redshift information currently derived in the zCOSMOS-bright project (Lilly et al. 2007, 2009) – a major spectroscopic redshift survey of the galaxies in the COSMOS field (Scoville et al. 2007). The final zCOSMOS-bright is designed to yield a spectroscopic information about ~20,000 galaxies with $I_{AB}<22.5$ across 1.7 deg$^2$ of the COSMOS field. With high success rate in measuring redshifts (close to 100% at $0.5 < z < 0.8$), good velocity accuracy (about 110 kms$^{-1}$), an anticipated high average final sampling rate across the field (~ 70%) and thanks to its highly continual angular completeness up to small scales (with no "fiber collisions" and with multiple coverage of the same area) it is well suited for the study of close kinematic pairs of galaxies.

The analysis is based on the data and the catalogues derived from the observations in the first two observing seasons. These observations yielded spectra for 10,509 galaxies. We refer to this spectroscopic sample as the "10k sample". More details on the design of the zCOSMOS and the 10k sample can be found in Lilly et al. (2007) and Lilly et al. (2009).

### 2.2 Selection criteria

We use the confidence system described in Lilly et al. (2009). For analysis in this paper we are using only redshifts with confidence classes 4.x, 3.x, 2.5, 2.4, 9.5, and 1.5. In addition so-called "secondary" targets with confidence classes as mentioned earlier are being used. These are spectroscopically targetable galaxies (i.e. satisfying the selection criteria of the survey) that fall serendipitously into the slit of another, "primary", target. Analysis of repeat observations indicates that the adopted sample is on average 99% reliable. It is considered unlikely that the very small number of incorrect redshifts will produce spurious pairs.

In an effort to work with a uniform sample of galaxies, we apply a rest-frame *B*-band absolute magnitude cut with an evolving limit of $M_B = -19.64 - 1.36\,z$ (here using $h = 0.7$) as in Kampczyk et al. (2007). This is at best an approximation to achieving a uniform volume-limited sample with redshift, but is better than making no attempt to correct for the luminosity evolution of individual galaxies.

### 2.3   Sample and its completeness

The overall redshift completeness of the galaxies selected as above, with absolute magnitude cut $M_{B,AB} < -19.64 - 1.36z$ is high, about 90%, and roughly constant over the redshift range of the sample (see Fig. 1). Photo-z estimates of the remaining objects suggest that they follow a very similar overall redshift distribution.

The use of a slit spectrograph – VIMOS (Le Fèvre et al., 2003) with multiple passes over the field (up to eight) means that the angular completeness is high down to small separations. In fact, the measurement of redshifts for so-called "secondary" targets lead to a slightly *higher* angular completeness on scales smaller than about 5 arcsec. The angular completeness is shown in Fig. 2.

### 2.4   Pair selection criteria and the close pair sample

Using secure redshifts and the aforementioned absolute magnitude cut yields a sample of 3667 galaxies up to $z \sim 1$ which can be used for finding close kinematic pair galaxies. We will refer to this (approximately) volume-limited sample of galaxies as the "overall" or "global" sample (we use both terms synonymously).

The criteria for the galaxies to be in a close kinematic pairs are as follows: We require a velocity difference $dv < 500$ kms$^{-1}$ (comfortably larger than the velocity accuracy of 110 kms$^{-1}$) and a projected proper (physical) separation $dr$ within some bound. We consider $dr < 100\ h^{-1}$kpc i.e. 140 kpc for our chosen $h = 0.7$, and also look at samples selected to have $dr < 30\ h^{-1}$kpc and $50\ h^{-1}$kpc. Additionally we require that the difference of the $M_B$ absolute magnitudes of the galaxies in a pair should not be larger than 1.5 mag., equivalent to a multiplicative factor of four, so that the galaxies in a pair can be regarded as a potential major merger system. These selection criteria yield 153 close kinematic pairs up to redshift $z = 1$ (see Fig. 3).

Detailed statistics of this pair sample as a function of redshift as well as details of completeness corrections can be found in our other work on close kinematic pairs in the zCOSMOS 10k sample - see de Ravel et al. (2011).

### 2.5   Environmental measures in zCOSMOS

One of the major goals of the zCOSMOS survey is to study environmental aspects of galaxy evolution up to $z \sim 1$. To achieve this goal, the three-dimensional density field has been reconstructed utilizing both spectroscopic (10k) and photometric redshifts (30k) for the full sample of galaxies up to $I_{AB} < 22.5$ in Kovač et al. (2010a). This reconstruction is based on the ZADE approach, which adjusts the redshift likelihood functions for the galaxies for which only photometric redshifts are available. Extensive tests on the COSMOS mock catalogues (Kitzbichler & White 2008) based on the Millennium simulation (Springel et al. 2005) demonstrate the robustness of this method in improving the reconstruction of the three-dimensional density field. Using an estimate of the mean density, Kovač et al. (2010a) then computed over-densities throughout the volume. A number of different density estimates have been derived with typical errors on log (1+δ) between 0.1 - 0.15 over the wide range of over-densities (see Kovač et al. 2010a for details). We will find it convenient below to consider the four quartiles of the distribution of over-densities traced by the overall galaxy population, ranging from D1 (lowest density) to D4 (highest density).

Knobel et al. (2009) have generated a group catalogue from the zCOSMOS 10k sample. This contains 800 groups with observed group richness (i.e. number of spectroscopically confirmed members) of $R \geq 2$, and 286 with $R \geq 3$. These are identified by combining both a friends of friends algorithm, and a Voronoi tessellation approach. The purity and completeness of this group catalogue, as assessed from mock COSMOS catalogues (Kitzbichler & White 2007) is quite high for a sample at these redshifts. The reader is referred to Knobel et al. for a more detailed discussion of this catalogue. All but a handful of our "pairs" at low redshift satisfy the group selection criteria and therefore appear in the group catalogue.

## 2.6 Morphological classification in COSMOS

We use the structural parameters and morphological classification of the COSMOS galaxies based on ZEST classification derived by Scarlata et al. (2007). The Zurich Estimator of Structural Types (ZEST) classification scheme is based on the principal component analysis of five non-parametric diagnostics, i.e. asymmetry A, concentration C, Gini coefficient G, second order moment of the brightest 20% of galaxy pixels $M_{20}$, ellipticity ϵ and sersic fits.

It should be noted, that these parameters are derived from the COSMOS HST I – band ACS images (Koekemoer et al. 2007), which were "cleaned" of nearby companions. The morphologies thereby derived should therefore reflect the properties of the individual galaxies in the close kinematic pairs *and should not be trivially biased by the presence of the nearby companion in the pair*, unless there is some true astrophysical interaction.

## 2.7 SFR and sSFR based on OII measurements

Spectra obtained in zCOSMOS-bright give measures of the emission line fluxes for the OII line in the redshift range of $0.5 < z < 0.9$. These in turn may be used to obtain star-formation rates, SFR - see Maier et al. (2009) for details. In addition specific star-formation rates, sSFR, defined as *SFR/m*, where *m* denotes the stellar mass of a galaxy, have been derived as in Maier et al. (2009). The sSFR is especially useful, since for the bulk of star-forming galaxies at a given redshift, the sSFR varies only weakly with mass over 2 orders of magnitude in stellar mass (Elbaz et al. 2007, Salim et al. 2007, Karim et al. 2011) and is apparently independent of environment (Peng et al. 2010) at least to $z \sim 1$. Comparison with the stellar masses of Maier et al. (2009) with those derived by Bolzonella et al. (2009) using the entire COSMOS optical to infrared photometry shows a good agreement with a statistical scatter of around 0.13 dex per galaxy.

## 3. Environmental dependence of the close kinematic pair fraction

## 3.1 Large scale environments of the close kinematic pair galaxies

In examining the environments of galaxy pairs, there are a number of subtleties to be considered in the choice of environmental density estimator. These are related to how the density is estimated and the sampling aperture.

First, many of the zCOSMOS environmental analysis have been based on counting galaxies within an adaptive aperture based on $N^{th}$ nearest neighbor. Clearly an environmental measure defined in this way will change as soon as a given pair merges into a single new galaxy, potentially leading to biases in the density distributions of pairs and single galaxies. This problem can be completely solved by using a stellar mass-weighted density that is computed on a fixed aperture that is much larger than the separations of the pairs. We have therefore adopted this approach in this paper, using 3 Mpc (comoving) as the radius, velocity offset of 1000 kms$^{-1}$ centered on the redshift of galaxy or a grid point and volume-limited tracers.

A second potential problem concerns the fact that some "pairs" will simply be chance alignments of two galaxies that are in reality separated by much more than 500 kpc. These galaxies will not merge for a considerable amount of time, if ever. Their impact on the merger rate may in principle be discounted by including a long timescale tail in the distribution of merger timescales (Kitzbichler & White 2008). The number of these "interlopers" will increase as the local projected density of galaxies increases. For instance, in a gravitationally bound group some galaxies will have small projected separations simply because of some coincidental alignment.

We correct for this potential effect by estimating the number of spurious alignments, using the actual density field and actual spectroscopic redshifts. We place 5 million times, one additional galaxy at a random location in the data and then see what fraction of these added galaxies form a "spurious pair" with one of the real galaxies in the sample, as a function of the over-density field at that location. This estimates the fraction of real galaxies that have a spurious companion simply because of projection effects (see Fig. 4), and we can therefore subtract this fraction from the observed fraction. As expected this correction increases both with *dr* and with environmental density. For the highest over-density quartile this correction decreases the fraction of pair galaxies by 23% for the pairs with *dr* < 100 h$^{-1}$kpc, 13% for those with *dr* < 50 h$^{-1}$kpc and 7% with *dr* < 30 h$^{-1}$kpc. This correction is significantly lower for the other quartiles and is essentially negligible for below the median of the density field (i.e. D1 and D2).

We can then compute pair-fractions, which are defined as the number of galaxies in pairs divided by the total number of objects in a given sample. In the simple case where all pairs are isolated pairs and not multiple systems (which is generally the case) this will be proportional to exactly twice the number of pairs. Fig. 5. shows the derived pair fractions in each of four quartile bins of over-density, for 3 different pair projected separations: *dr* < 100 *h*$^{-1}$kpc, *dr* < 50 *h*$^{-1}$kpc and *dr* < 30 *h*$^{-1}$kpc. The solid line symbols have been corrected for the projection bias, the uncorrected data are shown with dotted line symbols.

It can be seen that the corrected pair fractions increase with overdensity. The lowest quartile (defined by the global galaxy sample) of the over-densities contains only 10% of the pair galaxies, and the fraction of pair galaxies in the highest over-density quartile is typically 2-3 times higher than in the lowest quartile for all of the *dr* bins.

This clear environmental dependence of pair fractions will translate into a faster build-up of mass in denser environments via merging, assuming that the fraction of close pairs that will merge in a given timescale will not depend on environment. We examine this assumption in the next Section. Since the galaxy population as a whole varies with environment, this environmental dependence of the pair fraction also *requires that we must choose carefully the "parent" population of potential progenitors in examining the properties of the pair galaxies so as to match the environments.* This is explored further in Section 4.

## 3.2   Close kinematic pairs and merging in Millennium mock catalogues

In order to verify whether the environmental dependence of merging that we have found in zCOSMOS bright sample exists also in simulations, we use six pencil beam mock catalogues named Kitzbichler2006abcdef (Kitzbichler & White 2007) of a deep field of 1.4 times 1.4 square degrees based on the Millennium Simulation (Springel et al. 2005). First, we identify in cones in the redshift range 0.6 - 0.8 close kinematic pairs of galaxies with the same criteria as used for the observations. This includes the selection of a volume limited sample with an absolute magnitude cut $M_{B,AB} < -19.64 - 1.36 z$, selecting pairs with a velocity difference $dv < 500$ kms$^{-1}$ and a projected proper (physical) distance $dr$. As earlier we consider samples with $dr < 100$ $h^{-1}$kpc, $dr < 50$ $h^{-1}$kpc and $dr < 30$ $h^{-1}$kpc. We require that the absolute magnitudes of galaxies in a pair not differ by more than 1.5 mag.

In all 6 mocks (a, b, c, d, e, f) we also calculate the overdensities in a consistent way as for the zCOSMOS data. Calculated densities are mass-weighted in a fixed cylindrical aperture of 3 Mpc (comoving) radius and velocity offset of 1000 kms$^{-1}$, and use the same set of volume-limited tracers as above. Since they are mock catalogues there are no photo-$z$ objects. Based on the density distribution of the volume-limited sample of galaxies, we split the sample into four quartiles and use these when studying environmental dependencies in close kinematic pairs of galaxies.

By checking the survival of galaxies between different simulation snapshots, we trace whether the galaxies in pairs will merge by the end of simulation at $z = 0$. Fig. 6 shows the cumulative redshift distribution of merger events in close kinematic pairs chosen in a redshift range $0.6 < z < 0.8$, so that the fraction of galaxies that do merge by $z = 0$ is indicated by the values of $z = 0$. At fixed $dr$, the pairs in overdense and underdense regions that do merge have largely similar distributions of their merger redshift, but, for a given $dr$, the fraction of galaxies that will *not* merge in denser environments is higher then in the underdense regions.

We interpret this difference as the effect of the chance-projections that we identified earlier. Fig. 7 shows the fractions of galaxies in pairs for different $dr$, that will merge by redshift $z = 0$ as a function of density quartiles. Those fractions, especially for the sample with the largest projected separations $dr < 100$ $h^{-1}$kpc, decrease towards higher density. However, when we apply the correction for the "projection pairs", which we have derived in section 3.1, the trend with density vanishes. We therefore conclude that the fraction of galaxies at a given $dr$, that will merge by $z = 0$ is independent of the environment they reside in, as long as the observed pair fractions are corrected for the random, density-dependent, "false" associations caused by projection. The environmental dependence of pair fraction, suitably corrected, should therefore translate to an equivalent dependence on merger rate.

The mock catalogues show a broadly similar density-dependence in the pair fraction as the data. This is shown on Fig. 8. The fraction of galaxies in pairs in the mocks are plotted as open circles, the fraction of galaxies that were in pairs and merged by $z = 0$ in the simulation are shown with crosses. Both fractions show a clear trend with the large-scale environments that they reside in. This dependence is stronger in mocks than in our observational data, i.e. a factor of about seven between D1 and D4 as opposed to 2-3. Some of this may be due to noise in the zCOSMOS density field washing out the underlying effect. It is also possible that astrophysical effects (e.g. boosting of star-formation rates) which may be stronger in low-density environments (see below) are present in the data but not in the mock catalogues. For both reasons, our observationally derived over-density dependence of the pair fractions should be regarded as a lower limit.

### 3.3    Close kinematic pairs in zCOSMOS 10k groups

We use the 10k zCOSMOS group catalog from Knobel et al. (2009) to examine the dependence of the close kinematic pair abundance as a function of group richness. Almost by definition, virtually all pairs will be found in the catalogue of "groups", which extends down to groups with two members, since our pair selection criteria exceeds (except at very low redshifts) the minimum linking length of the group-finding algorithm. Fig. 9 examines, as a function of group richness, the fraction of group galaxies that are selected to be close kinematic pairs according to our criteria. The richness represents the observed richness of the group, based on the number of spectroscopically confirmed members. Groups with an observed richness equal 2 are the most abundant ones in the group catalog and, similarly, nearly half of pair galaxies are found in 2 member groups.

The pair fractions of galaxies in different richness groups are shown in Fig. 9, compared also with the overall pair fraction regardless of environment. Since essentially all of the galaxies in close kinematic pairs are in groups with $R \geq 2$, the pair fraction in groups is about four times higher than overall, the same factor by which the galaxy sample is larger than the group galaxy sample. There is a rather small dependence of pair fraction with the observed group richness, and within the error bars it can be assumed to be flat. There is a possible increase of the pair fractions in the richest groups ($R > 7$), but this should be treated with caution, both because of its limited statistical significance and, as discussed in Section 3.1 above, we would expect random contamination to increase with density and projected separation.

### 3.4    Summary of environmental effects

To summarize this Section 3, we find that the incidence of pairs, once corrected for projection effects, is about 2-3 times higher in the highest density quartile (D4) than in the lowest density quartile (D1). This translates also to the same difference in the merging rate, since the distributions of timescales are thought to be similar. While almost all pairs are, by definition, found in groups, we do not find a strong dependence on group richness.
Over 70% of the galaxies seen in close kinematic pairs with $dr < 50h^{-1}$kpc at $0.5 < z < 1.0$ will have merged with their companion by $z = 0$. This fraction is largely independent of environment, once the correction for the effect of chance-projections is applied.

### 4.    Morphological types of galaxies in close kinematic pairs

### 4.1    Morphology of galaxies in pairs

We first look at the morphological fractions of paired and non-paired galaxies, i.e. the fraction of galaxies that have a particular morphological type. These are shown in Fig. 10, at different projected separations (vertically) and as a function of redshift (left to right in the left-hand plots). The different morphological types are color-coded according to the ZEST classification: red – type 1 (spheroid dominated), blue – type 2 (disc-dominated), and black – type 3 (irregular). The narrow panel shows the average morphological fractions throughout the redshift range $0.2 < z < 1$. The histograms on the right show the *relative* morphological fractions of close kinematic pair galaxies for $dr < 100\ h^{-1}$kpc and $dr < 50 h^{-1}$kpc, normalized with respect to the morphological fraction of the overall sample of galaxies that is shown in the top panel.

It can be seen on this plot that the morphological mix of close kinematic pairs share some common trends with that of the underlying overall sample of galaxies. Disc galaxies dominate all three samples (overall galaxies and the two sets of close pairs) at all redshifts. The fraction of spheroids increases towards lower redshifts, while the fraction of galaxies that are classified as irregulars decreases.

The pair samples, however, show some significant differences in their morphological mix compared to the overall sample. In particular, the fractions of both the spheroid galaxies and of the irregular galaxies are both boosted in the close kinematic pairs relative to the overall sample. However, we show below that this is due to the fact established above that the close kinematic pair galaxies reside preferentially in denser environments, where the spheroid fraction is higher, reflecting the morphology-density relation that is seen in the zCOSMOS 10k sample extending out to redshift $z = 1$ (e.g. Tasca et al. 2009).

The boosting of the irregulars clearly increases to the smaller separations and reaches a relative 50% excess for the sample with $dr < 50\ h^{-1}$kpc compared to the overall sample. This increase with decreasing radius probably reflects a real morphological change due to the closer separation of the galaxies in these closest pairs.

The environmental dependence of the pair fraction, and the environment-dependence of the galaxy morphological mix, mean that the morphological properties of the pairs can only be compared with carefully matched samples of non-pair galaxies. This is done in the next Section.

## 4.2  Modeling morphology fractions with Monte Carlo models

To investigate what drives the morphological mix in the close kinematic pairs, we have performed a series of Monte Carlo simulations. In each of these, we select, from the parent population of galaxies, a new progenitor population with the same number of pairs (for both *dr* separations) that are chosen to match the observed pair samples in four key properties. In each case, 1000 Monte Carlo realizations are made of each of these four samples. The four comparison samples are constrained to have:

Z:   exactly the same redshift distribution as the pair galaxies
ZO:  the same redshift distribution *and* over-density distribution as the pair galaxies
ZG:  the same redshift distribution and the galaxies belong to groups in the group catalog
ZG3: the same redshift distribution and the galaxies are in the group catalog with an observed richness $R \geq 3$ or higher.

As noted above, the pair fraction is observed to be more or less constant with group richness, and so it might be thought that ZG was the ideal comparison sample and that ZG3 might in a sense "overshoot" in environmental matching. We include ZG3 because of possible concerns that the close pairs may be more "group-like" than the looser more typical R=2 groups of Knobel et al., which may be more contaminated with chance projections of field galaxies.

Figure 11 shows the relative morphological fractions of the galaxies in the observed close kinematic pairs, at different separations, compared with those that would be expected from each of the above 4 "parent" populations. The color-filled areas show the 68% confidence interval derived from the Monte Carlo simulations.

It can be seen in Fig. 11 that the simplest model "Z" under predicts the spheroid fractions by about 30 – 40%, at about a 2-sigma level. This fraction is slightly better reproduced by the model that takes into account the over-density distribution of the pair galaxies – "ZO", but the fractions of spheroids in the close kinematic pairs with projected separations $dr < 100$ $h^{-1}$kpc and $dr < 50$ $h^{-1}$kpc are only correctly reproduced by the models involving group galaxies – i.e. "ZG" & "ZG3" respectively.

All the Monte Carlo models under predict the numbers of irregular galaxies at all separations. At separations $dr < 50$ $h^{-1}$kpc, the irregulars are over-represented in the observed pair sample by about 50% relative to the model "Z". This excess increases to more than 75% in comparison to those models that should best match the parent population, and which correctly accounts for the spheroid fractions – i.e. "ZG3".

The fact that the spheroid fraction in pairs is well-matched to the parent population in this environment (and therefore also that the overall "late-type" fraction disks plus irregulars is well-matched) suggests that the excess of irregulars in the pairs originates primarily from morphological disturbance of disk galaxies. The values in Fig. 11 suggest that about 10% of the disk galaxies change their appearance due to the interaction in the pair sufficiently to be classified as irregulars at separations of $dr < 50$ $h^{-1}$kpc. The fact that the spheroid fraction is not perturbed presumably reflects the fact that interactions are more likely to produce dramatic distortions of morphology in dynamically cold systems.

## 4.3   Galaxy-galaxy morphological combinations in the close kinematic pairs

Another way of analyzing the morphological segregation is to look at the galaxy-galaxy morphologies in the close kinematic pairs (Fig. 12), by which we mean the fraction of galaxies that have a particular combination of morphologies of the *two* components, using exactly the same methodology of constructing "random" pairs from the differently-matched progenitor samples.

Fig. 12 shows the fractions of the observed pairs exhibiting various combinations of morphologies, and, on the right, how these fractions compare with those expected from the ZG and ZG3 control samples. As would be expected, most pairs contain at least one late-type galaxy, and any subsequent merger would be expected to be "wet" in the popular parlance. Although the fraction of spheroid-spheroid pairs is higher than expected from the most basic control sample, this excess is found to largely disappear when the control sample is progressively better matched to the environments of the real pairs.

As before, the strongest effect is a significant excess in the number of irregular-irregular pairs, by about a factor up to 8. Specifically, the irregular-irregular fractions for $dr < 100$ $h^{-1}$kpc is four times higher relative fraction and for $dr < 50$ $h^{-1}$kpc it is six times that expected from the ZG control sample, and eight times for ZG3. This can be compared with the factor of only about 1.2-1.7 excess in the overall number of irregular galaxies in the sample identified in the previous section. In other words, the irregular-irregular excess is larger than would be expected if the "extra" irregulars were randomly distributed around the pairs, suggesting that irregularity is mutually triggered: if one galaxy is made irregular in the pair, the other one is more likely to have been similarly perturbed.

In the underlying (non-pair) sample, some 16% of irregular galaxies have an irregular companion in a pair. In pairs with $dr < 100$ $h^{-1}$kpc, fully a half of the irregular galaxies have an irregular companion, and in fact at $dr < 50$ $h^{-1}$kpc, only about 40% of their companions are *not* irregular. The fraction of irregular-irregular pairs increases with redshift in all three samples.

### 4.4 Summary of morphological effects

To summarize the above discussion: Any interpretation of the morphological mix of galaxies observed to be in close kinematic pairs must take into account the environments in which these pairs are typically found, in order to correctly remove the environmental dependence of general galaxy properties. When this is done, it is found that the main effect is an increase in the fraction of irregular morphologies by a modest factor – up to 75% in the fraction of irregulars, or equivalently, a conversion of about 10% of the disk galaxies into irregulars, with the fraction of spheroids in pairs closely matching what would be expected. The fraction of irregular-irregular pairs is further boosted, suggesting that if one galaxy is irregular, the other is much more likely to be irregular, and in fact 60% of the irregulars in pairs are in irregular-irregular pairs, despite comprising only 15% of the sample as a whole.

### 5. Star-formation activity in close kinematic pair galaxies

### 5.1 The induced star-formation in close kinematic pair galaxies

In this Section we compare the distribution of specific star formation rates (sSFR) in the close kinematic pair galaxies with those in the underlying parent samples, as constructed using the matching techniques described above.

In Fig. 13, we show the distributions of log(sSFR) in the pairs, at $dr < 100$, 50, 30 $h^{-1}$kpc compared to the ZO, ZG, and ZG3 parent samples. A significant enhancement in sSFR is seen at the smaller projected separations. This excess appears to involve a modest enhancement in sSFR of a significant fraction of the galaxies, although the sSFR of the most passive (lowest sSFR) galaxies appears to show little change. Fig. 14 shows the boost in the average log(sSFR) as a function of pair separation compared with the different comparison samples. The boost increases as we progressively match the samples and reaches between 2 and 4 for the smallest separations $dr < 30$ $h^{-1}$kpc. This mirrors a similar effect seen in our analysis of AGN fractions of paired galaxies (Silverman et al. 2011).

### 5.2 The contribution of induced star-formation in pairs to the global star-formation rate

We can use the fraction of galaxies in pairs, together with the sSFR boost computed in the previous section, to estimate the fraction of the total star-formation seen in the galaxy population which is associated with the star-formation in close kinematic pairs, and also what fraction of this is due to the enhancement discussed in the previous section.

When corrected for incompleteness in the spectroscopic sampling, the fraction of galaxies in close kinematic pairs above $M_{B,AB} < -19.64 - 1.36\,z$ in the redshift range $0.5 < z < 0.9$ is $0.06 \pm 0.01$ for pairs with separations $dr < 30\ h^{-1}$kpc - see de Ravel (2011). This is the fraction of galaxies that would be in pairs *if* all galaxies had been observed spectroscopically.

The excess in the sSFR at $dr < 30\ h^{-1}$kpc, relative to the simplest Model Z, is $1.9 \pm 0.6$. This implies that close kinematic pairs are contributing about $11\% \pm 4\%$ of the star-formation seen in the *B*-selected ($M_{B,AB} < -19.64 - 1.36\,z$) galaxy sample, although only about a half of this, i.e. 5%, is due to the "excess" star-formation associated with the interaction itself. We conclude that interactions in the early phases of a merger represent a detectable but small contribution to the overall star-formation budget of the Universe at these redshifts.

In the 10k sample, 26% of galaxies are in groups. This fraction would be expected to increase to about 33% if all galaxies had been observed spectroscopically. Since all pair galaxies are in a group and we expect the pair fraction in a fully-sampled survey to be about 2.9 times higher, the contribution of paired galaxies with $dr < 30\ h^{-1}$kpc to the star-formation budget in groups will be correspondingly higher by a factor of about 2.3, i.e. increasing to about 30% (of which about 15% is "excess").

## 5.3   Post-starburst galaxies in close kinematic pairs

Galaxies that show signs of a starburst and subsequent cessation of star formation in their spectra were first identified by Dressler and Gunn (1983) on the basis of the combination of strong Balmer absorption lines, indicating a large stellar population of age $10^8$-$10^9$ years, and a lack of significant emission lines indicating little ongoing star-formation. This class of galaxies could be a transition phase between the blue and red sequences.

While many such galaxies have been found in galaxy surveys, it is still not clear what physical mechanisms are responsible for this phenomenon. There are several proposed scenarios. Ram pressure gas stripping, harassment or strangulation are proposed as an efficient mechanisms for quenching star-formation activity in an over-dense environments (Gunn & Gott 1972; Cayatte et al. 1994; Larson et al. 1980; Moore et al. 1996, 1998; Balogh & Morris 2000). Internally in the galaxies themselves, star-formation activity could be suppressed by strong AGN/SN feedback (Springel et al. 2005; Hopkins et al. 2007). Galaxy interactions and merging can trigger star formation (Toomre & Toomre 1972; Barnes & Hernquist 1992; Naab & Burkert 2003). Strong starburst present in such interactions could also cause a rapid exhaustion of the fuel supply resulting in ceasing of star-formation activity.

Because of the possible importance of galaxy interactions for creating a post-starburst (PSB) phase, it is instructive to check whether they can be found in samples of close kinematic pairs and what would be their abundances. The VVDS survey has identified PSB galaxies (Wild et al. 2008) and one of them turned out to be in a sample of 36 close kinematic pairs (de Ravel et al. 2009).

Post-starburst "k+a" galaxies have been identified in the 10k zCOSMOS bright sample based on the measurements of the Balmer break and EW[OII] by Vergani et al. (2010). In the redshift range of 0.48 < *z* < 1.0, 35 post-starburst galaxies have been identified within our volume limited sample of galaxies used to identify close kinematic pairs in the present work. Out of those 35 galaxies, five (14%) are in a close kinematic pairs with *dr* < 100 kpc. The fraction of post-starburst galaxies in pairs therefore appears to be twice higher (2.4% ± 1%) than in the underlying sample of galaxies (1.2% ± 0.2%), although the statistical significance of this increase is marginal. It is also interesting to notice, that Vergani et al. (2010) finds "k+a" galaxies preferentially in a rich environments, although not exclusively. This is similar to what we have inferred about the environments of the close kinematic pairs.

Interestingly, there is some evidence that groups contain an excess of PSB galaxies (1.66% ± 0.5%) over the field (1.2 ± 0.2%). However if the contribution of paired galaxies in groups is taken out, the remaining group galaxies show no excess in PSB fractions in comparison to the global sample (1.29% ± 0.5%). This could indicate that, in the group environment, the PSB phase is mostly triggered in close kinematic pair galaxies. If this result holds up with better statistical significance in larger samples, then it will show that the interactions associated with the early stages of mergers are able to quench as well as boost star-formation.

## 6. Environmental dependence of the effects of interactions and possible biases

In the Sections 4 and 5 above, we have showed that the morphological mix of paired galaxies can be well explained by a combination of simple environmental effects, i.e. taking into account the morphological mix in equivalent environments, plus interactions in close pairs that evidently transform some spirals into irregulars. Similarly, the distributions of sSFRs in pairs were well matched to comparison samples at large separations, but are evidently boosted (at least in previously star-forming galaxies) in close pairs, suggestive of induced star formation during interaction.

In this section we show that, perhaps surprisingly, these manifestations of interaction-induced activity are preferentially seen in regions of lower overall galaxy density, *even though the majority of pairs are found in over-dense regions*.

Fig. 15 shows the morphological fractions of the close kinematic pair galaxies and of the overall sample in four quartiles of over-density. Both in the overall sample and in the pair samples with different separations there is the well-known trend of increasing spheroid fraction and decreasing irregular fraction with density. These trends are however stronger in the pair population than in the overall sample mostly because of the marked increase in irregulars in the close pairs. It can be seen that there are actually no spheroids in the D1 (lowest density) quartile in the pairs with separations *dr* < 50 $h^{-1}$kpc, and the dominant morphological type is the irregulars with 55% of galaxies, nearly 3 times higher than in the overall sample. In the highest quartile of over-density distribution, the excess of irregulars has largely vanished along with a marked decrease in the overall irregular fraction.

Non-parametric diagnostics like asymmetry *A*, concentration *C*, and Gini coefficient *G* have frequently been used to characterize the morphologies of merger candidates. Here we focus on the asymmetry *A*, which has been found to be the best discriminator of merger candidates as showed in Kampczyk et al. (2007).

Fig. 16 shows the distribution of *A* for different samples of close kinematic pairs split by separation *dr*, by environment and by redshift. In each panel two or more reference lines are given representing either the parent galaxy population as observed in zCOSMOS, or the distribution of asymmetries for 115 *visually-selected* merger candidates identified in the COSMOS field by Kampczyk et al. (2007) in a redshift range $0.7 < z < 0.8$.

Several things are visible in this Figure. First, the distribution of $A$ in the pairs is always skewed to higher $A$, especially at smaller $dr$, but it never reaches the asymmetries of the visually identified mergers. The fraction of galaxies in pairs with asymmetries larger than 0.2 is 12%, 17% & 21% for the separations of $< 100$ $h^{-1}$kpc (blue), $< 50$ $h^{-1}$kpc (green) & $< 30$ $h^{-1}$kpc (red) respectively while for the underlying sample this fraction is only 7% (black line). For comparison, 33% of visually identified mergers have $A > 0.2$. This is also shown in Table 1. Second, it is apparent that the increase in $A$ is greater in the lower density environments. This can be partly explained by the increased incidence of late type galaxies (disk and irregular) in low density environments, but it also requires that these later type galaxies be more asymmetric in the pairs in the lower density regions than in the higher density ones. Finally, in the third panel, the increase in A in the lower density environments is more noticeable in the higher redshift pairs.

The boost in the specific star formation rates in pairs also seems to be environmentally dependent. At $dr < 30$ $h^{-1}$kpc, the excess reaches to factors of 2 - 4 compared with the best matched comparison samples. In Fig. 17, this excess is shown to be largest in the lowest density parts of the sample, while being modest – less then a factor of 2 – and only significant at about 1 sigma level in over-dense regions. This cannot be simply explained by the change of morphological mix observed in different environments. The environmental dependence weakens, when only star-forming galaxies with $\log[sSFR/Gyr^{-1}] > -1$ are selected, but does not vanish.

The fact that both morphologically, and in terms of the sSFR, the interactions in lower density regions seem to have a stronger effect contrasts with the evidence that sSFRs (of star-forming galaxies) seem to be largely independent of over-density at given epoch, as shown in Peng et al. (2010). We could speculate that perhaps the interactions affect larger scale gas reservoirs that do not change the steady-state star-formation rates of normal galaxies, and that these could be larger in lower density regions. Further investigation, using larger samples, is needed to explore this.

Our analysis here suggests that two of the most important consequences of interaction - morphological disturbances and boosted star-formation - may be more important in lower density environments, where relatively few such pairs are found. The environmental dependence on the pair fraction derived in paragraph 4.3, coupled with the evidence from simulations that the timescales for a given pair to merge are largely independent of environment, suggests that the majority of future merger progenitors reside in over-dense environments, where they show little evidence for either enhanced sSFR or asymmetries. This could give rise to at least two observational biases toward the lower density systems: first, for luminosity-selected samples, boosted star-formation may brighten galaxies into the samples; second, morphologically selected samples of mergers may also be biased towards lower density regions. We noted above that mock catalogues have a stronger environmental dependence on merging rate (about six times higher in D4 than D1) than the observed sample. This bias could explain some of this.

Given the evidence for higher gas fractions at high redshift (e.g. Tacconi et al. 2010), it is quite possible that higher values of the evolution parameter $m$ could result. High redshift late type bulgeless galaxies are more likely to have bar instabilities that would drive gas to their central regions causing boosts of star formation (Mihos & Hernquist 1996; Di Matteo et al. 2007). Those galaxies are expected to be more prone to induced star-formation than the bulge dominated ones. Observed build up of bulges in disc galaxies between $z = 1$ and $z = 0$ (Oesch et al. 2009) gradually stabilizes them against violent instabilities (Mihos 2000) and makes them more resistant to induced star-formation events like those observed in the close kinematic pairs.

Close kinematic pair studies based on the precise redshift information, at least for the bright galaxies, have consistently given lower values of "*m*" and a shallower redshift evolution (Neuschaefer et al. 1997; Lin et al. 2004; Lin et al. 2008; de Ravel et al. 2011) when compared to studies based on morphological classifications (Le Fèvre et al. 2000; Conselice et al. 2003; López-Sanjuan et al. 2009; Bridge et al. 2010). Even at low redshifts, where recognition of a interaction-induced morphological disturbance is easier (Kampczyk et al. 2007), there seem to be little overlap between samples of the close kinematic pairs and those based on highly asymmetrical or disturbed systems (de Propris 2005).

## 7. Summary

In this paper we have analyzed the properties of the close kinematic pairs of galaxies selected from the spectroscopic zCOSMOS 10k bright catalog in a volume limited sample spanning over the redshift range $0.2 < z < 1.0$. Close kinematic pairs are defined to have velocities within 500 kms$^{-1}$ and projected separations *dr* up to 100 $h^{-1}$kpc. Using the multi-wavelength information available from the COSMOS and a reconstructed 3D density field we find that:

(a) Close kinematic pairs of galaxies preferentially reside in overdense environments. The fraction of galaxies in close pairs with *dr* < 100 $h^{-1}$kpc is 1.8 times higher in the highest over-density quartile, D4, than the lowest one, D1, increasing to 3 times higher for pairs within 50 h$^{-1}$kpc. The majority of close pairs reside in over-dense environments. Examination of the Millennium mock catalogues suggests that the fraction of these systems that will merge (and the timescale for them so doing) is essentially independent of environment, so this variation in the pair fraction should translate to a variation in the merger rate.
(b) Almost all pairs are in groups (by definition), but there is no strong trend of pair fraction with group richness. It follows that the merging rate for galaxies in groups is higher than in the field (all galaxies) by the inverse of the fraction of galaxies in groups, which in our sample is 3.8.
(c) The morphologies of the close kinematic pair galaxies are not representative of the underlying global galaxy sample, but rather reflect the morphological mix of the richer environments that they reside in. The fraction of spheroidal types in the pairs is higher than for field galaxies, but is exactly the same as for the population of group galaxies. The fraction of irregulars in the close kinematic pairs is however significantly elevated, at the expense of disk galaxies, presumably because some of the latter are perturbed into an irregular appearance. This excess is especially pronounced when comparing with group galaxies, and increases as the pair separation decreases. Reflecting this effect, the distribution of asymmetries of galaxies in pairs is skewed towards higher values especially in the closest pairs *dr* < 30 $h^{-1}$kpc.
(d) The combinations of morphologies in individual close kinematic pairs also show an increase in the relative number of irregular – irregular pairs indicating that morphological disturbance in one galaxy is usually accompanied by a disturbance of the other. In fact, only ~ 40% of the irregular galaxies in close kinematic pair sample with $dr < 50\ h^{-1}$kpc do *not* have an irregular companion. For $dr < 50\ h^{-1}$kpc there is an excess of irregular-irregular pairs by a factor of about 4 to 8. Nevertheless, disk-disk pairs are the dominant type in our sample, and spheroid–spheroid pairs, which would be expected to be future "dry" merger systems, are the least abundant morphological combination, contributing on average only ~5% of the close kinematic pair systems in the redshift range $0.2 < z < 1$. This fraction probably increases towards low redshift due to the increase of the early-type population.

(e) Galaxies in close kinematic pairs show an enhancement of SFR and sSFR that increases with decreasing projected separation. The excess in the sSFR at $dr < 30$ kpc, is about a factor of two ($1.9 \pm 0.6$). Coupled with the overall pair fraction (corrected for spatial sampling), this implies that close kinematic pairs are contributing about 10% of the star-formation seen in our volume limited galaxy sample, about a half of which can be attributed to "excess" star-formation associated with the interaction.

(f) Post-starburst galaxies are two times more abundant among close kinematic pair galaxies than in the underlying sample. This, together with evidence for triggered star formation in the close kinematic pairs, indicates that, at least in some cases, galaxy interactions in pairs can also be responsible for closing down star-formation.

(g) Interestingly, the effects of the interactions on the morphologies and star-formation rates of individual galaxies are most apparent in pair galaxies at close separations in low-density environments, where the boost in sSFR can reach 2-4 times. Pair galaxies in above-average over-dense environments, which comprise the majority of the close kinematic pair systems, do not show strong signatures in their distributions of asymmetries, sSFR, or irregular fractions when compared with suitable parent galaxy samples. This environmental difference, which is not apparent in the sSFR of normal non-interacting star-forming galaxies, may reflect the action of gas reservoirs that do not control (directly) steady-state star-formation.

(h) Studies based on identifying galaxies or pairs with clear signs of interactions may therefore be biased to lower density environments, sampling a minority of kinematic pairs. The severity of this bias may be redshift dependent. The same may also be true of flux-limited samples if excess star-formation boosts the brightness of galaxies into samples. We might expect these effects to produce steeper redshift evolution of disturbed object fractions than in a studies based on close kinematic pair systems, and suggest that this may be the cause of the wide discrepancies in the literature about the redshift dependence of the merger rate.

Overall, our study demonstrates that close kinematic pairs preferentially reside in high density environments, and that this must be taken into consideration in interpreting the properties of the galaxies involved in comparison with control samples of non-interacting galaxies. Our results demonstrate that interactions, which may represent the earlier phases of a galaxy merger, have a detectable, but generally not a dominant effect on galaxy evolution. Despite the care with which we have attempted to isolate selection effects, our results hint at further potential biases with environment and redshift that must be carefully controlled in the future. The forthcoming 20k zCOSMOS bright sample will double the number of redshifts and is expected to improve statistics of the close kinematic pairs by a factor of ~ 3.

Acknowledgements

The zCOSMOS survey was undertaken at the ESO VLT as Large Program 175.A-0839. We gratefully acknowledge the work of many individuals, not appearing as authors of this paper, whose work has enabled large surveys as COSMOS and zCOSMOS. The Millennium Simulation databases used in this paper and the web application providing online access to them were constructed as part of the activities of the German Astrophysical Virtual Observatory. This work has been supported in part by the Swiss National Science Foundation.

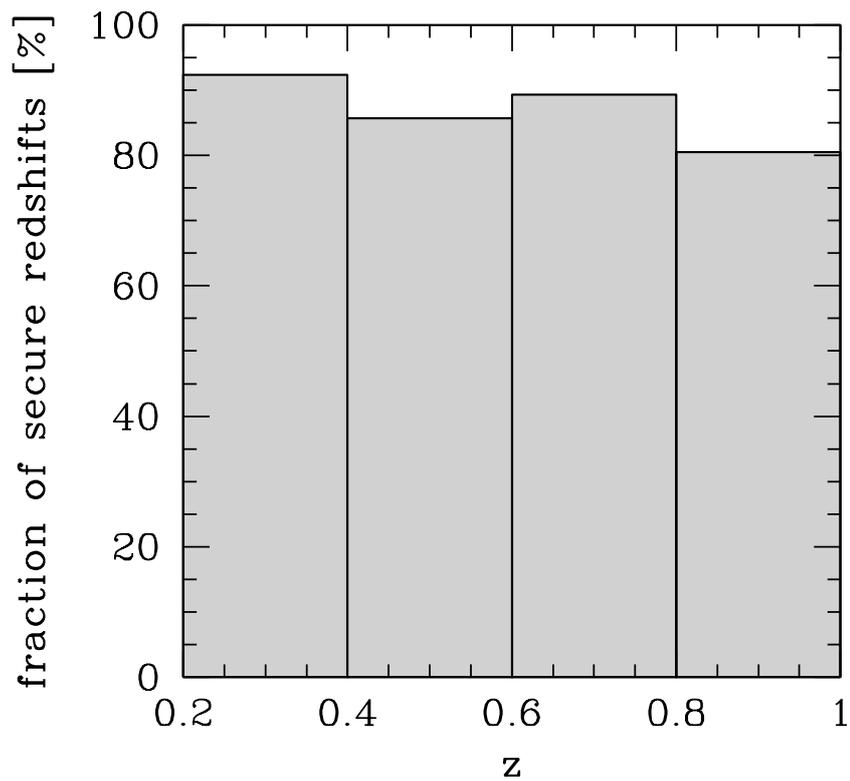

*Fig. 1: Redshift completeness of the zCOSMOS 10k subsample with secure redshifts in a volume limited sample defined with an absolute magnitude cut $M_B < -19.64 - 1.36\,z$.*

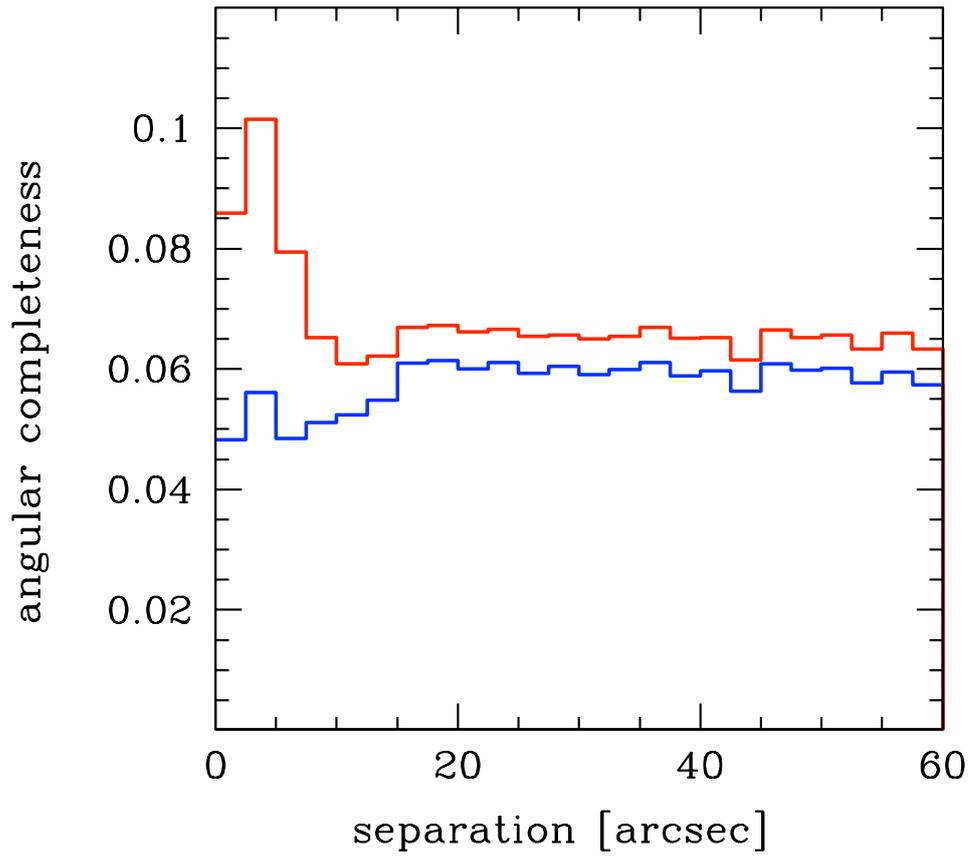

*Fig. 2: The angular completeness of the 10k sample defined as the fraction of projected pairs in which both members were observed, compared with the case if all galaxies had been observed spectroscopically. Blue: only primary spectroscopic targets. Red: together with the secondary detections used in this analysis.*

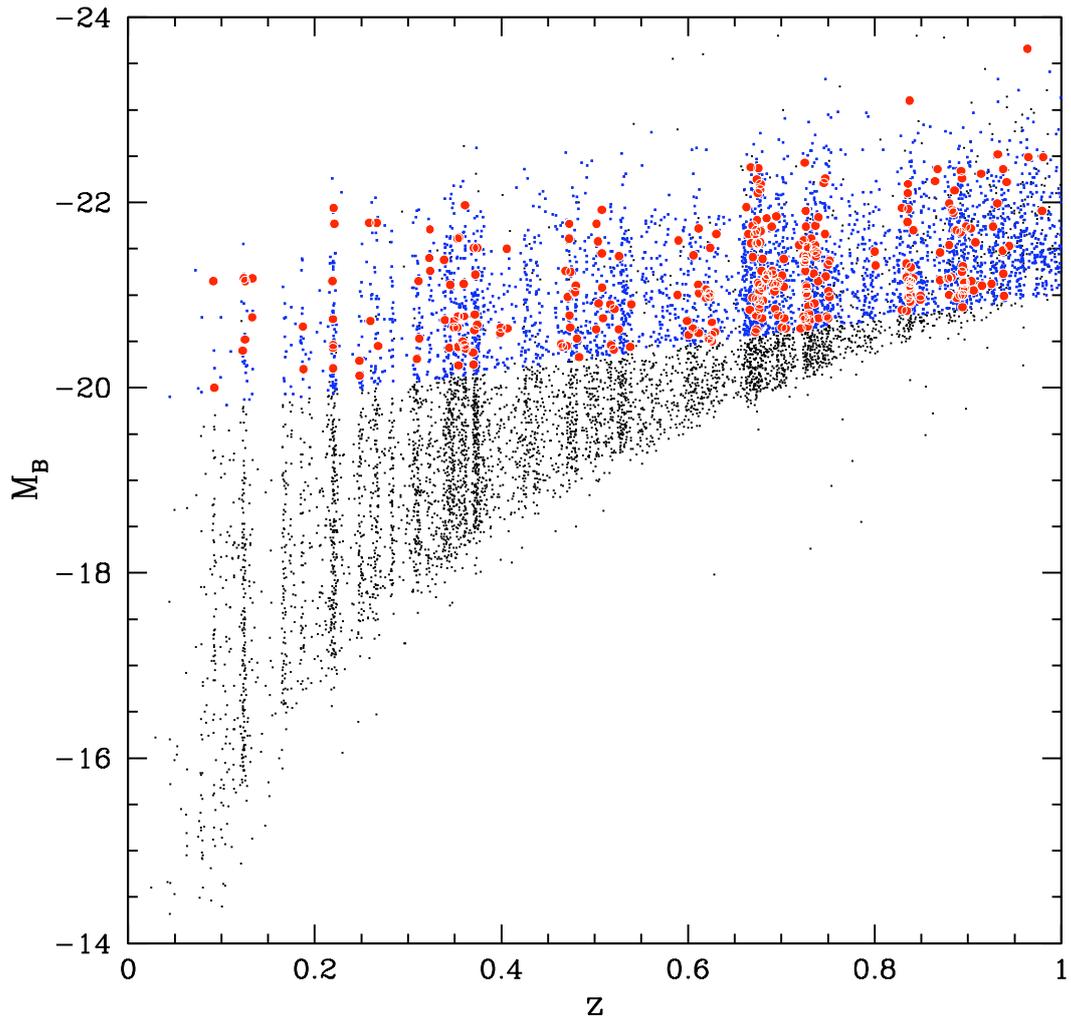

*Fig. 3: The rest frame B-band absolute magnitude as a function of redshift for the zCOSMOS 10k sample. The blue points represent galaxies with secure redshifts above the absolute magnitude cut ($M_B < -19.64 - 1.36\,z$) among which pair galaxies are highlighted with red dots.*

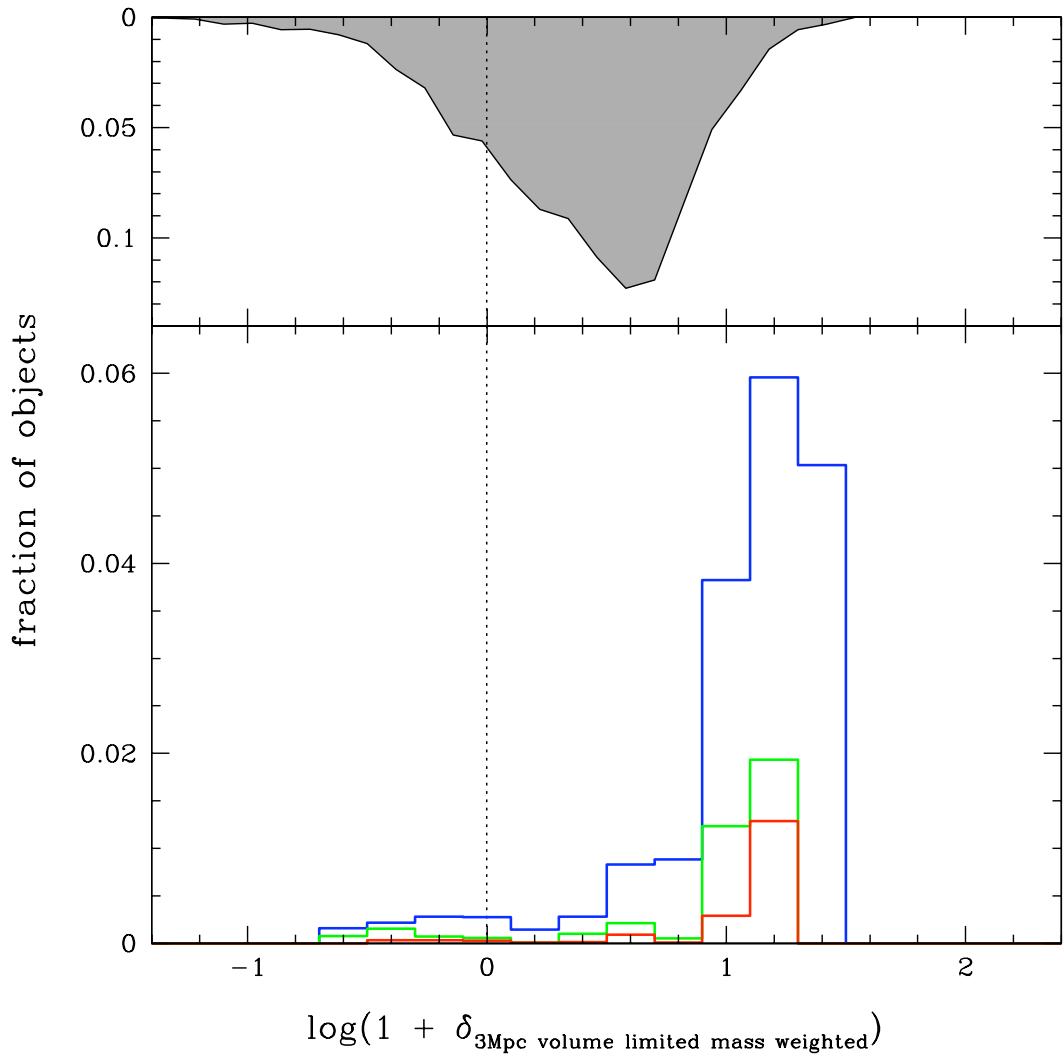

*Fig. 4: Lower panel: The fraction of galaxies in the 10k sample that would be associated as a close kinematic pair if an additional galaxy is randomly inserted into the density field. Close pairs with projected separations of dr < 100 $h^{-1}$kpc are in blue, dr < 50 $h^{-1}$kpc in green, and dr < 30 $h^{-1}$kpc in red. Top panel: the distribution of over-densities of objects in the 10k sample.*

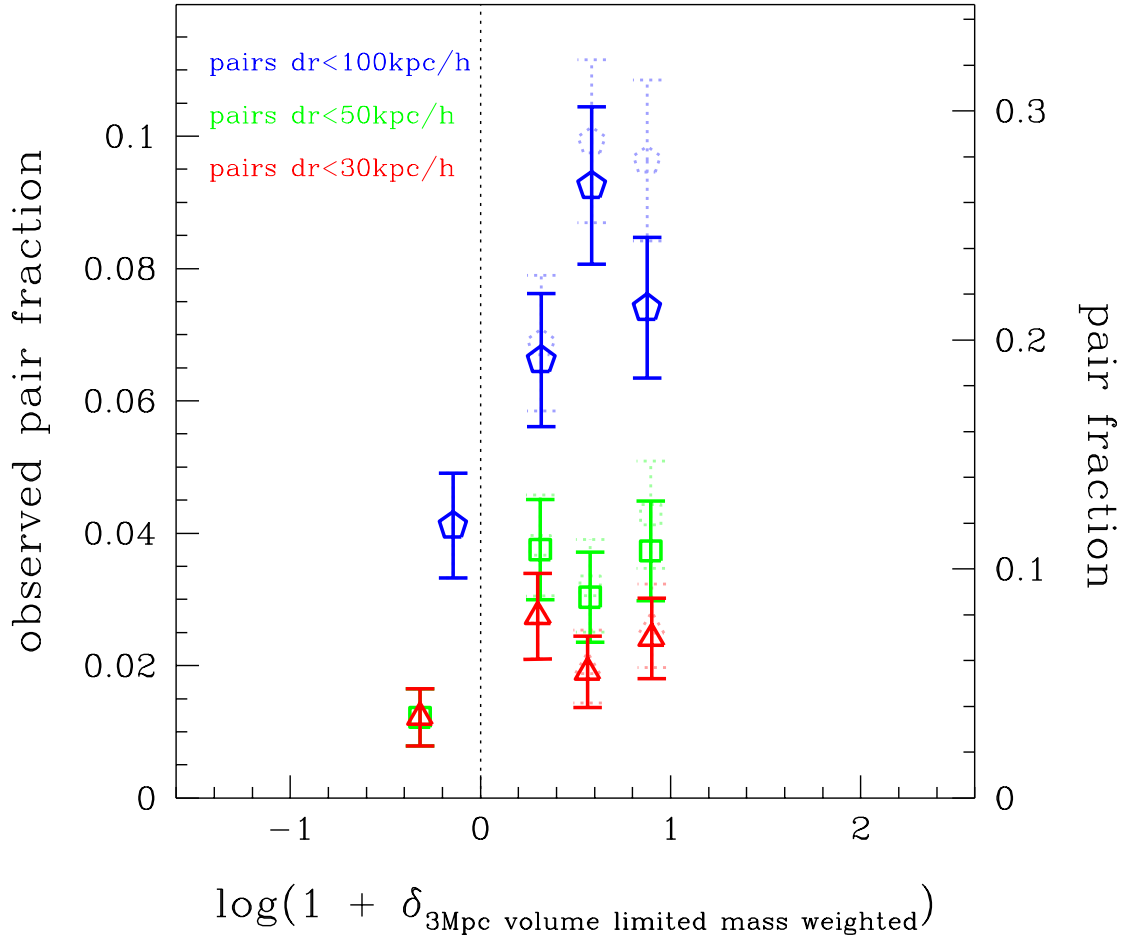

*Fig. 5: The fraction of galaxies in close kinematic pair systems as a function of over-density, computed as a mass-weighted, 3 Mpc constant aperture calculated with volume limited tracers, dr < 100 h$^{-1}$kpc in blue, dr < 50 h$^{-1}$kpc in green, and dr < 30 h$^{-1}$kpc in red, divided into quartiles of the density distribution of the underlying galaxy population. Dotted lines represent observed pair fractions, while the solid ones are after applying a correction for random unrelated projections as discussed Section 3.1. The left vertical axis shows the fractions observed in the 10k sample, the right axis shows the fractions expected in a complete, volume limited sample in which all galaxies were observed spectroscopically.*

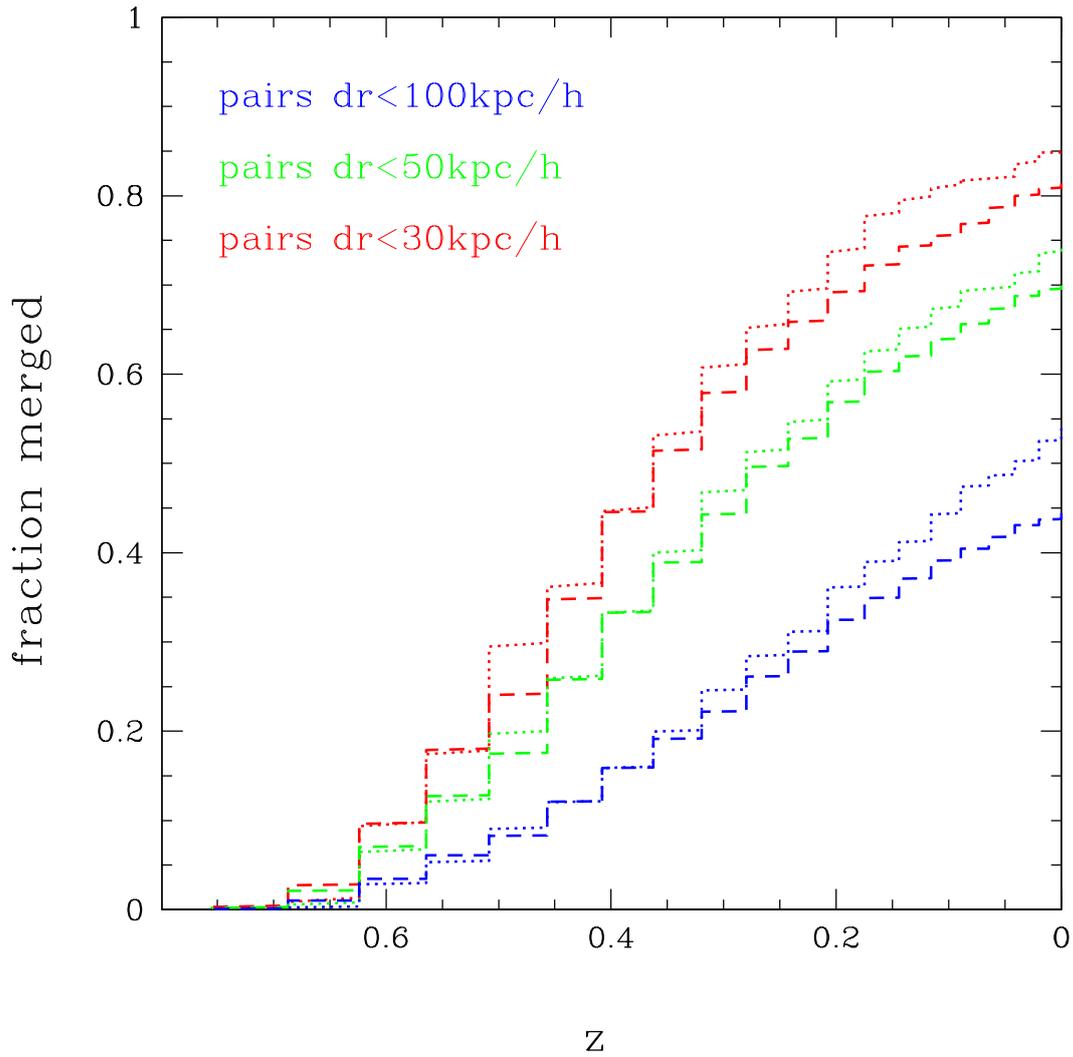

*Fig. 6: The cumulative redshift distribution of merger events in close kinematic pairs of galaxies selected in the redshift range 0.6 < z < 0.8 from the Kitzbichler 2006 mocks of the Millennium simulation. Galaxy pairs were chosen with the same absolute magnitude cut and major merger criterion as in our observed sample: dr < 100 $h^{-1}$kpc in blue, dr < 50 $h^{-1}$kpc in green, and dr < 30 $h^{-1}$kpc in red. Dashed lines indicate the distributions for pairs residing in environments denser than the median of an underlying density distribution, dotted – residing in a lower than the median. The fraction of galaxies that merged by z = 0 is indicated by the values at z = 0.*

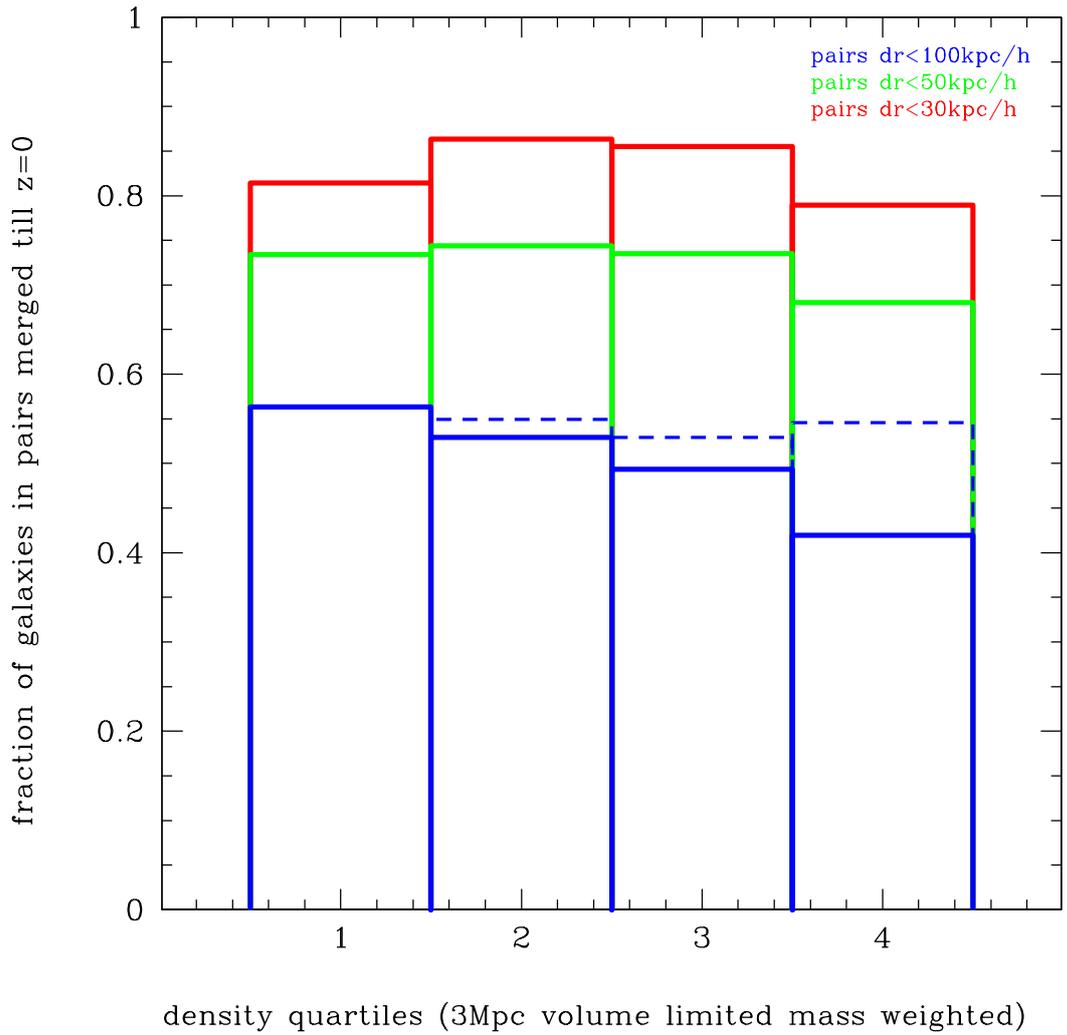

*Fig. 7: The fraction of galaxies in pairs at 0.6 < z < 0.8 that merged in the Millennium simulation by z = 0 as a function of environment: dr < 100 $h^{-1}$kpc in blue, dr <50 $h^{-1}$kpc in green, and dr <30 $h^{-1}$kpc in red. The dashed line indicates the fractions of galaxies in pairs with dr < 100 $h^{-1}$kpc that would merge till z = 0, after applying the correction for "spurious pairs" constructed in Section 3.1.*

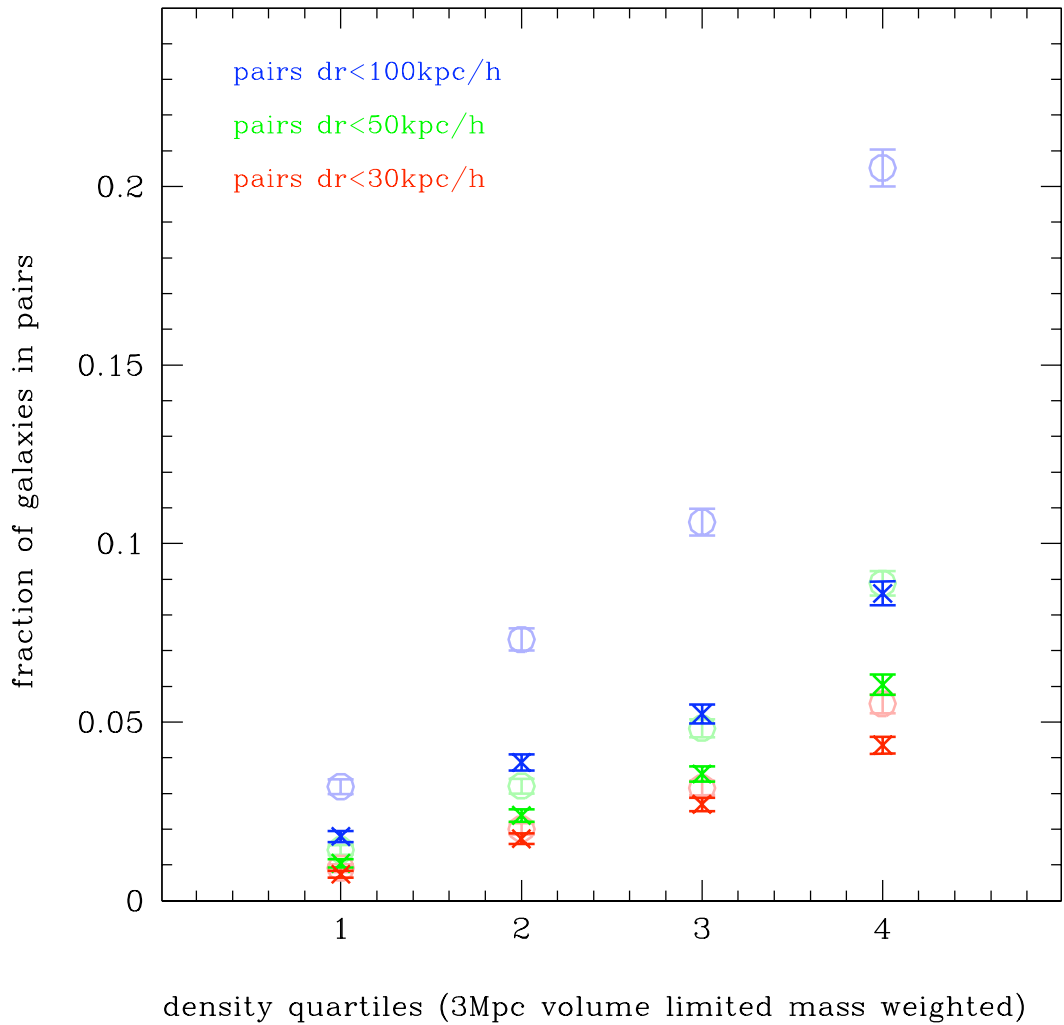

*Fig. 8: The fraction of galaxies in the close kinematic pair systems in the Kitzbichler2006abcdef Millennium mocks in the redshift range 0.6 < z < 0.8 as a function of environment: over-densities are calculated as in our zCOSMOS sample. These are mass weighted, 3 Mpc constant-aperture with volume limited tracers: $dr < 100\ h^{-1} kpc$ in blue, $dr < 50\ h^{-1} kpc$ in green, and $dr < 30\ h^{-1} kpc$ in red, divided into quartiles of an underlying galaxy population. The "observed" fractions are indicated with open circles, while the crosses show the fractions of galaxies that are identified at 0.6 < z < 0.8 in pairs and which merged in the simulation by z = 0.*

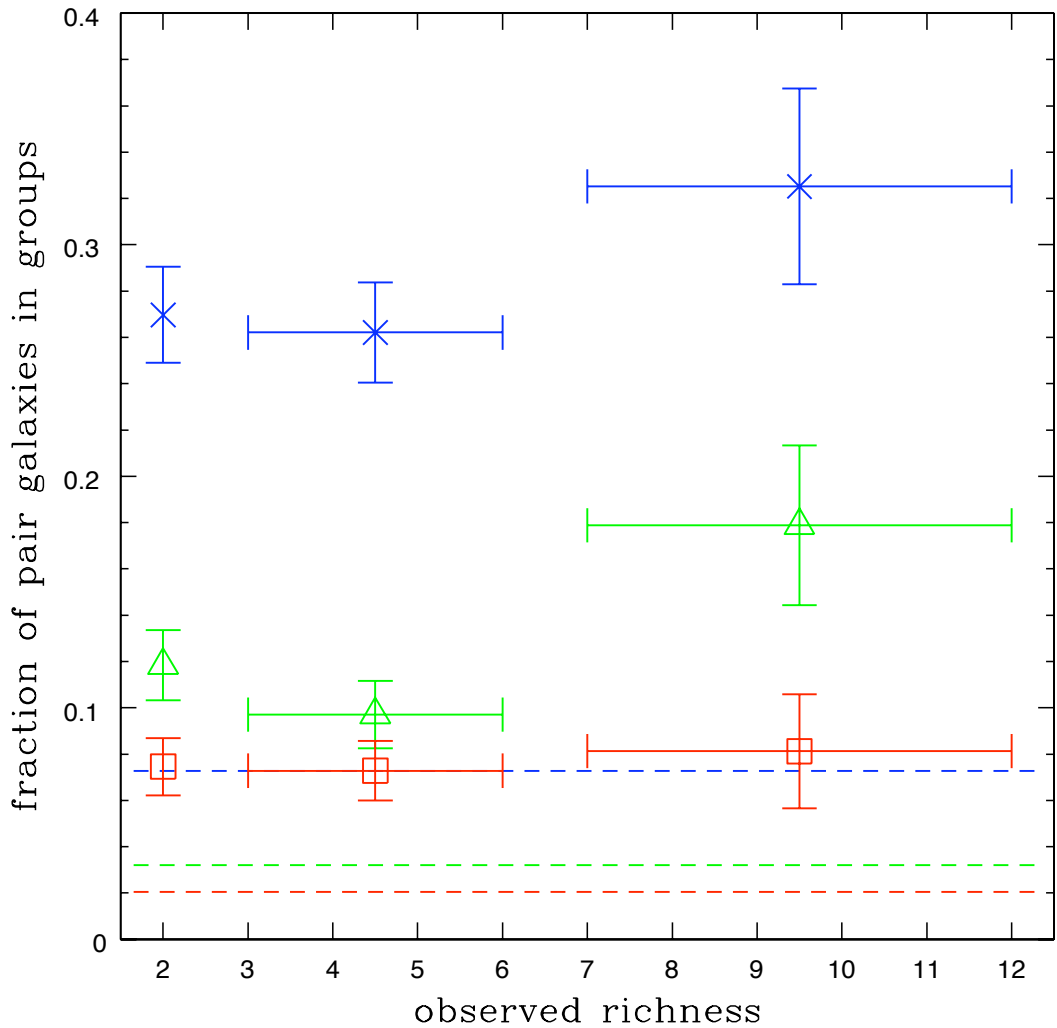

*Fig. 9: The fraction of the galaxies in groups that are in the close kinematic pair systems at a given observed group richness, with dr less than 100 $h^{-1}kpc$ (blue crosses), less than 50 $h^{-1}kpc$ (green triangles), less than 30 $h^{-1}kpc$ (red squares). Vertical error bars are the standard $1\sigma$ errors for the binomial distribution, while the horizontal ones show the binning. Dashed horizontal lines indicate the overall observed pair fraction calculated with respect to all of the galaxies in the volume limited zCOSMOS 10k sample, color-coded at the same projected distances.*

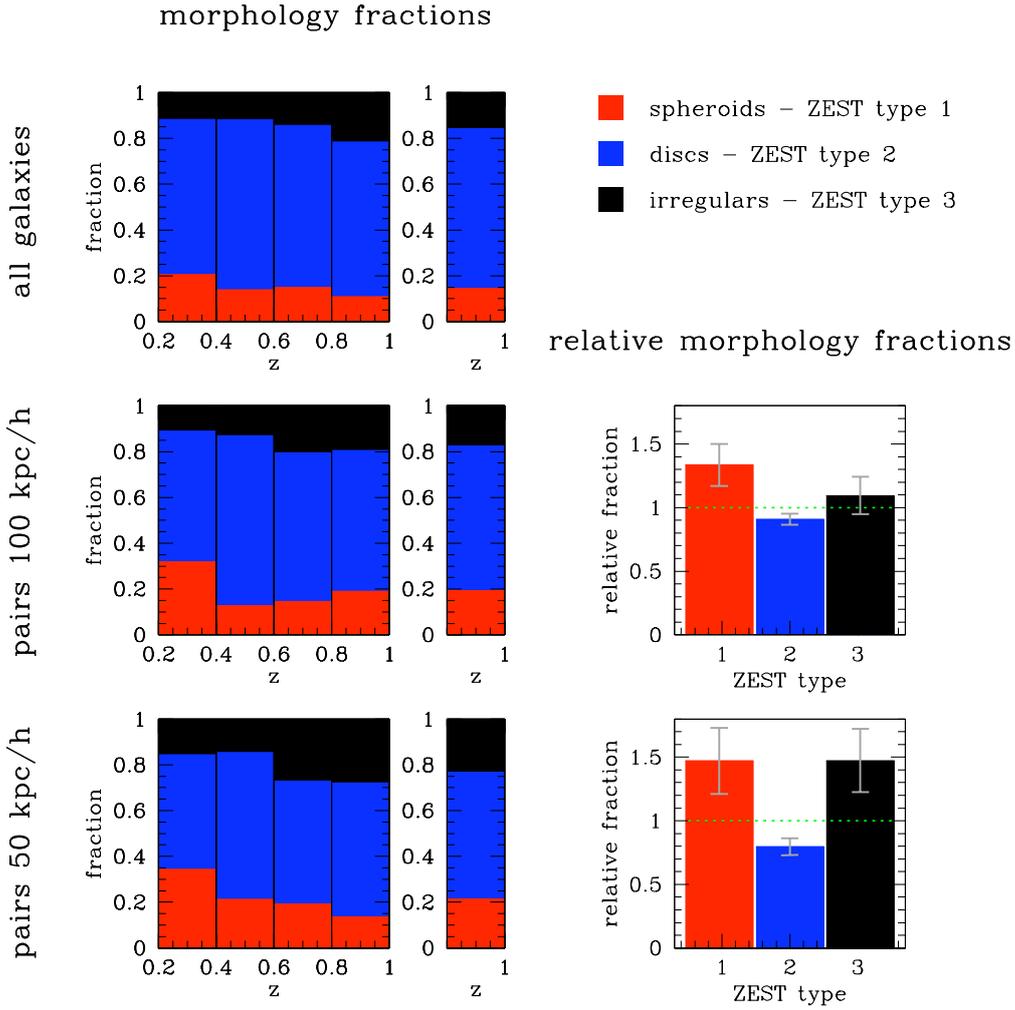

Fig. 10: Morphologies of the close kinematic pair galaxies. The top-left panel presents morphologies of all galaxies in the underlying sample (not matched in environment) in the different redshift bins. Different morphological types are color-coded according to the ZEST classification: red: type 1 (spheroid), blue: type 2 (disc), black: type 3 (irregular). Top right, the narrow panel shows the average morphological fraction through the redshift range $z = 0.2 - 1$. With the same convention, the two mid – left panels show the morphological fractions for close kinematic pair galaxies with $dr < 100\,h^{-1}kpc$, while the two bottom – left panels for close kinematic pair galaxies with $dr < 50\,h^{-1}kpc$. Panels on the right show the relative morphological fractions of the close kinematic pair galaxies with a projected separations $dr < 100\,h^{-1}kpc$ & $dr < 50\,h^{-1}kpc$ in respect to the underlying sample.

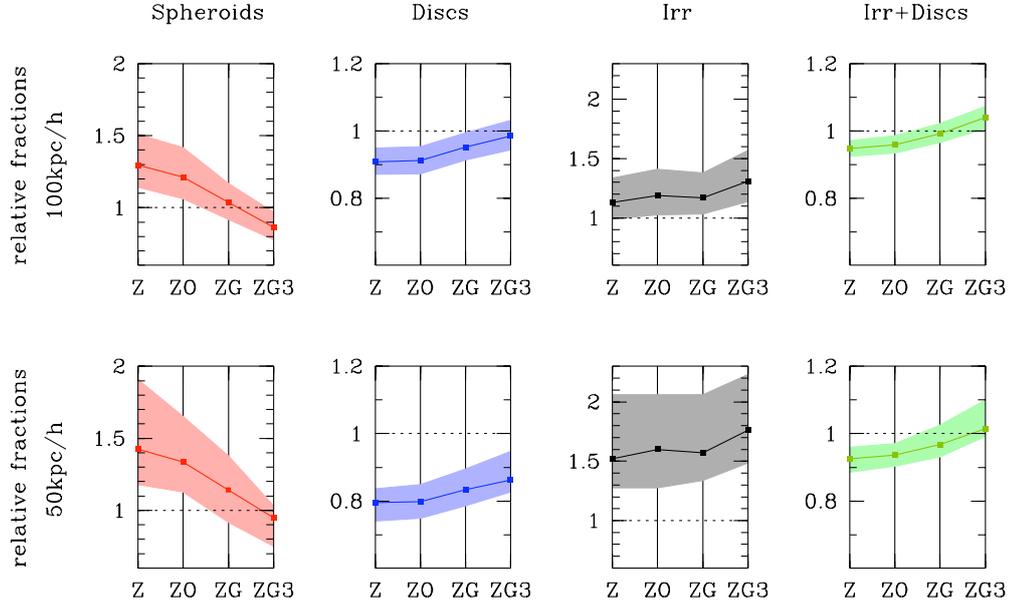

*Fig. 11: Relative morphology fractions of the close kinematic pair galaxies in respect to the morphological type abundances derived in 4 different models in the Monte Carlo simulations. Filled areas show the derived 68% confidence interval. The upper panels present the relative morphological fractions of the close kinematic pair galaxies with $dr < 100\ h^{-1}kpc$, while the lower ones are for the close kinematic pair galaxies with $dr < 50\ h^{-1}kpc$. The four models indicated on plots represent progressively better matching with the environments of the pair galaxies, and are: "Z" – galaxies drawn randomly from all of the galaxy sample to follow the redshift distribution as in the pair galaxies; "ZO" – the same redshift distribution and over-densities as in the pair galaxies; "ZG" – redshift distribution as in the pair galaxies and galaxies are in a group catalog; "ZG3" – redshift distribution as in pair galaxies and galaxies are in a group catalog with an observed richness of 3 or higher. The fractions of spheroids (or irregulars + discs) are best reproduced by ZG & ZG3 models for the close kinematic pairs with projected separations $dr < 100\ h^{-1}kpc$ & $dr < 50\ h^{-1}kpc$ respectively.*

*Fig. 12: Galaxy – galaxy morphological combinations of the close kinematic pair galaxies, color-coded according to the ZEST classification: black – type 3 & type 3 (irr – irr), green – type 3 & other, blue – type 2 & type 2 (disc-disc), yellow – type 1 & other, red – type1 & type 1 (spheroid-spheroid). The top left panel presents galaxy – galaxy morphologies constructed from all of the possible pairs of all the galaxies in a given redshift bin regardless their distance. The narrow panel on the right shows an average galaxy – galaxy morphology fractions in the redshift range z=0.2-1. With the same convention, the two mid–left panels present galaxy–galaxy morphological fractions for close kinematic pair galaxies with dr < 100 $h^{-1}$kpc, while the two bottom–left panels show close kinematic pair galaxies with dr < 50 $h^{-1}$kpc. The four panels on the right show the relative galaxy–galaxy morphological fractions of the close kinematic pair galaxies with a projected separations dr < 100 $h^{-1}$kpc & dr < 50 $h^{-1}$kpc in respect to the ones from the global sample and in respect to our preferred Monte Carlo models: ZG (for the pair sample with dr < 50 $h^{-1}$kpc) and ZG3 (for the pair sample with dr < 100 $h^{-1}$kpc).*

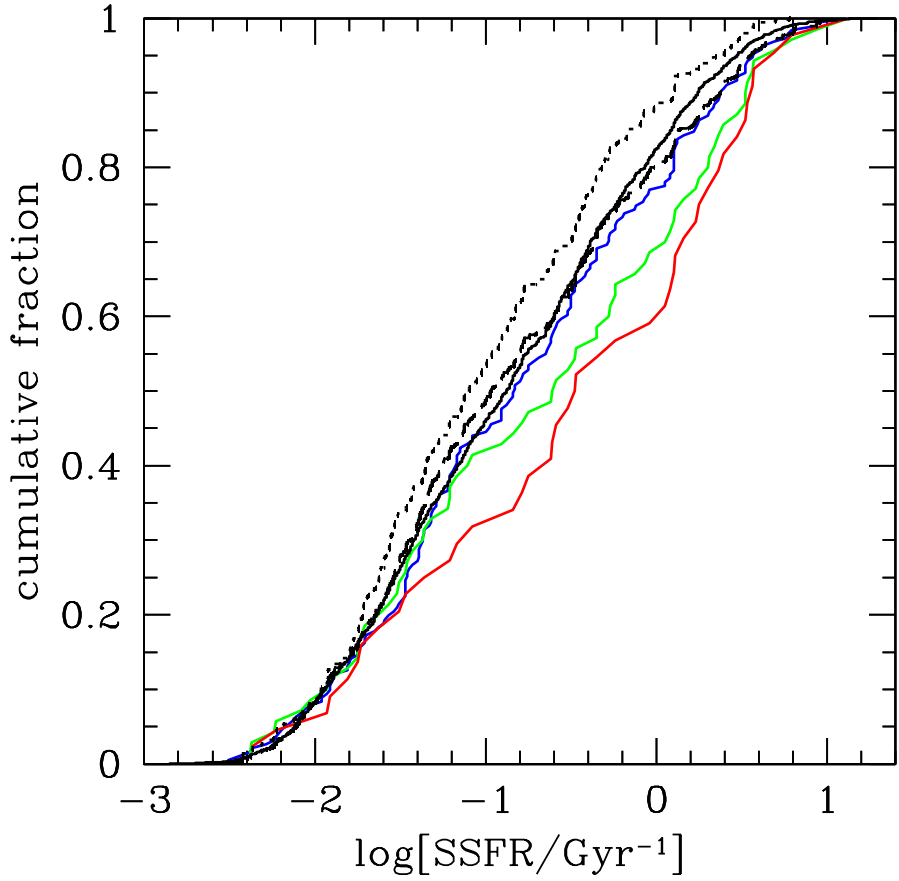

*Fig. 13: Cumulative distribution of sSFR based on [OII] 3727 measurements for Monte Carlo comparison samples in black "ZO" – redshift distribution and over-densities as in the pair galaxies (solid line); "ZG"- redshift distribution as in the pair galaxies and galaxies are in a group catalog (dashed line); "ZG3" – redshift distribution as in the pair galaxies and galaxies are in a group catalog with an observed richness of 3 or higher (dotted line). The sSFR of galaxies in pairs is indicated with colors: blue – with dr <100 $h^{-1}kpc$, green – with dr <50 $h^{-1}kpc$, red – with dr <30 $h^{-1}kpc$.*

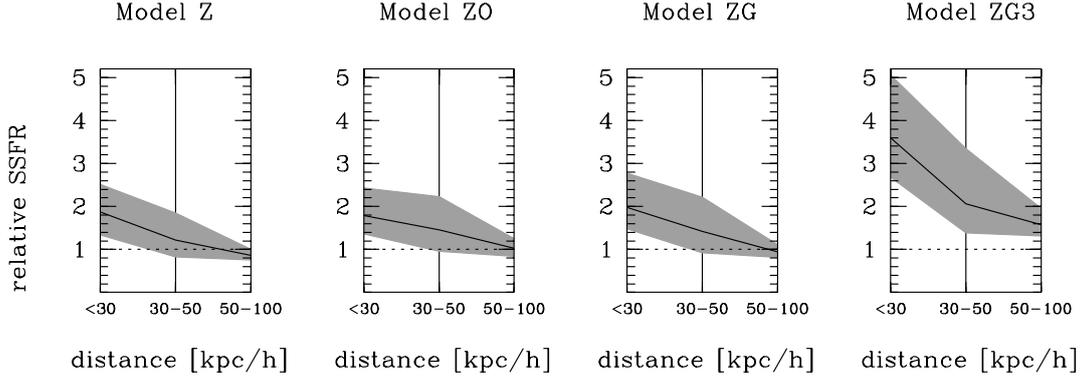

*Fig. 14: Relative average sSFR in the close kinematic pair galaxies with respect to the sSFR derived for 4 different models in Monte Carlo simulations as a function of projected separations in 3 bins: less than 30 $h^{-1}$kpc, 30 – 50 $h^{-1}$kpc and 50 – 100 $h^{-1}$kpc. Filled areas show the derived 68% confidence interval. The four models indicated on the plots are: "Z" – galaxies drawn randomly from all galaxy sample to follow redshift distribution as in pair galaxies; "ZO" – redshift distribution and overdensities as in the pair galaxies; "ZG"- redshift distribution as in the pair galaxies and galaxies are in a group catalog; "ZG3" – redshift distribution as in the pair galaxies and galaxies are in a group catalog with an observed richness of 3 or higher.*

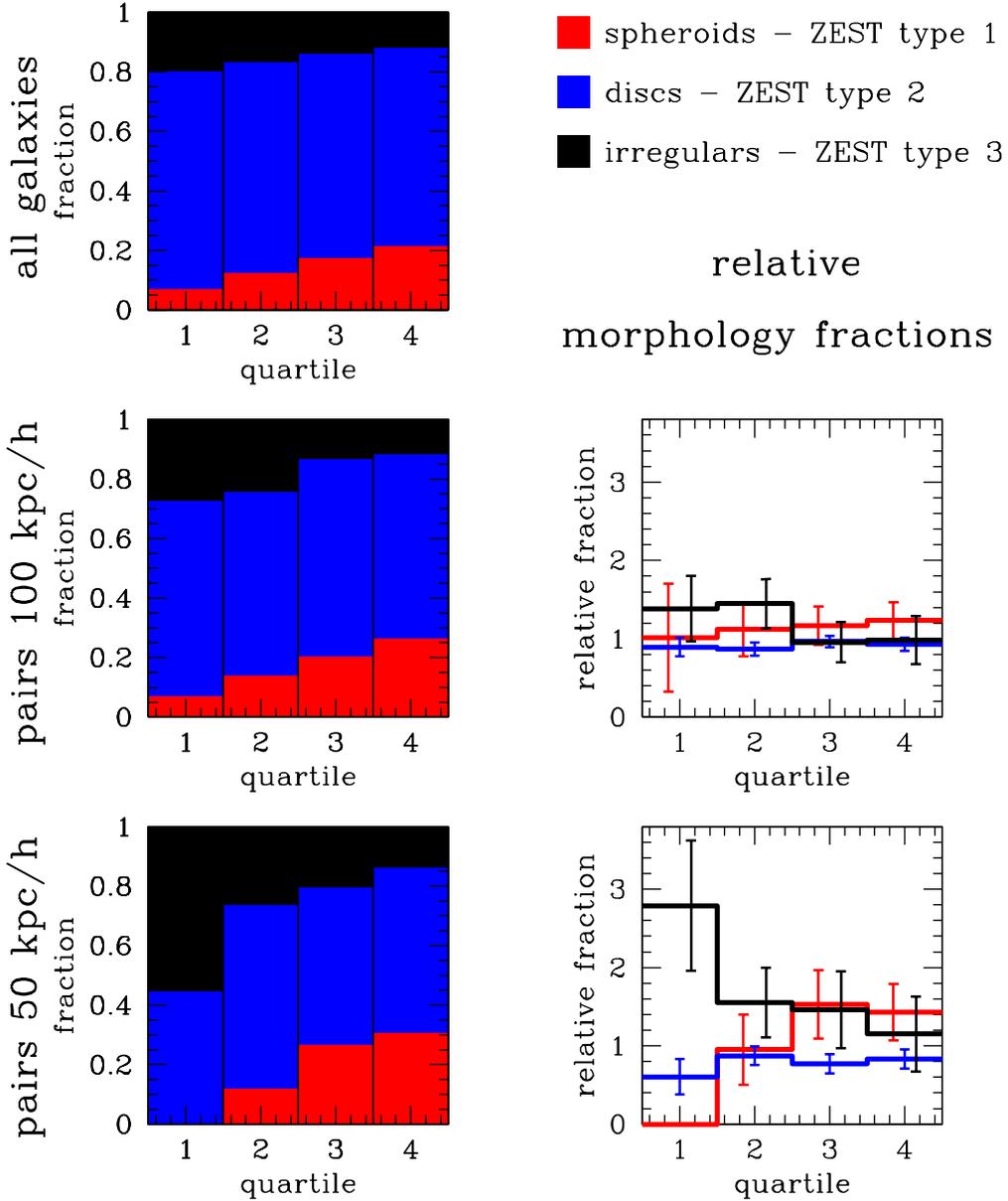

Fig. 15: Morphologies of the close kinematic pair galaxies as a function of overdensity quartiles. Top-left panel presents morphologies of all galaxies in an underlying sample in quartiles of the 3 Mpc constant aperture, mass weighted overdensities calculated with the volume limited tracers. Morphological types are color-coded according to the ZEST classification: red: type 1 (spheroid), blue: type 2 (disc), black: type 3 (irregular). With the same convention, the mid-left panel presents morphological fractions for the close kinematic pair galaxies with $dr < 100\ h^{-1}kpc$, while the bottom-left one shows the close kinematic pair galaxies with $dr < 50\ h^{-1}kpc$. Panels on the right show the relative morphological fractions of the close kinematic pair galaxies with a projected separations $dr < 100\ h^{-1}kpc$ (middle) and $dr < 50\ h^{-1}kpc$ (bottom) in respect to an underlying sample.

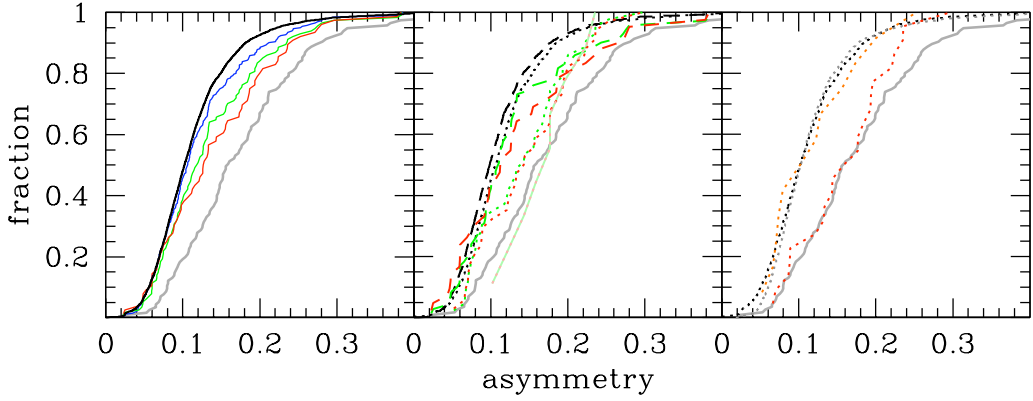

*Fig. 16: Cumulative distributions of asymmetries: on all panels the grey solid line represent the distribution of asymmetries for 115 visually selected merger candidates in the COSMOS at z =0.7 - 0.8 (Kampczyk et al. 2007). The left panel presents distributions for galaxies in the underlying sample (black), the close kinematic pair galaxies with projected separations less than 100 $h^{-1}$kpc (blue), less than 50 $h^{-1}$kpc (green) and less than 30 $h^{-1}$kpc (red). The central panel shows distributions with the same color scheme as before, but the distributions are divided according to the environments – dotted lines for the galaxies that reside in the lower than median over-density, dashed – above the median over-density. Dotted light green–light red line on the right side of the grey solid one represents the data for the lowest quartile (D1) of the over-density distribution for the close pairs with less than 50 $h^{-1}$kpc and less than 30 $h^{-1}$kpc (both samples coincide in this quartile). The right panel presents the result of splitting the underlying sample with lower than median over-densities into two redshift subsamples of an equal size – the high redshift one (black dotted line) and the low redshift one (grey dotted line). The close kinematic pair galaxies with projected separations less than 30 $h^{-1}$kpc were split around the same redshift into the high redshift bin (red dotted line) and the low redshift one (orange dotted line).*

|  | mean | median | SIQR |
|---|---|---|---|
| zCOSMOS ALL | 0.113 | 0.102 | 0.030 |
| pairs dr < 100 $h^{-1}$kpc | 0.122 | 0.106 | 0.036 |
| pairs dr < 50 $h^{-1}$kpc | 0.136 | 0.120 | 0.048 |
| pairs dr < 30 $h^{-1}$kpc | 0.143 | 0.128 | 0.056 |
| 115 COSMOS mergers | 0.179 | 0.159 | 0.056 |

Table 1: Asymmetries statistics for various pair, global and merger galaxy samples (see text for the details).

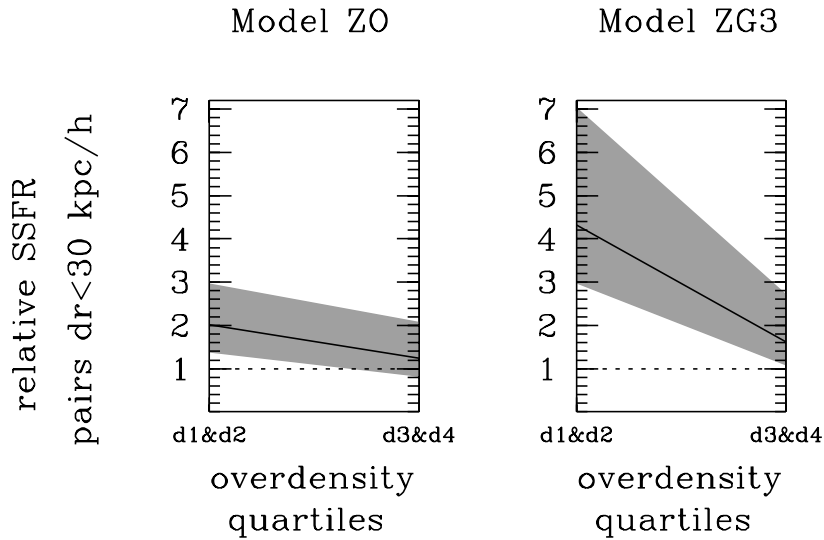

*Fig. 17: Relative average sSFR in the close kinematic pair galaxies with projected distances less than 30 $h^{-1}$kpc in respect to the sSFR derived for 2 different models in our Monte Carlo simulations as a function of environment. Galaxies in lower than median over-density quartiles D1 & D2 show much stronger enhancement in sSFR than the galaxies residing in higher than median over-density quartiles D3 & D4. The two models are: "ZO" – redshift distribution and over-densities as in the pair galaxies; "ZG3" – redshift distribution as in the pair galaxies and galaxies are in a group catalog with an observed richness of 3 or higher.*